\documentclass[twocolumn]{aastex631}

\usepackage{academicons}
\usepackage{amsmath}
\usepackage{hyperref}
\usepackage{natbib}
\usepackage{xcolor}
\usepackage{xspace}
\usepackage{subfigure}
\usepackage{threeparttable}

\shortauthors{Van Zandt et al.}

\graphicspath{{./}{Figures/}}

\newcommand{\hipAtrend}{\ensuremath{-1.2^{+0.2}_{-0.2}}\xspace}
\newcommand{\hipAcurv}{\ensuremath{\equiv 0}\xspace}
\newcommand{\hipAdmu}{\ensuremath{0.43^{+0.05}_{-0.05}}\xspace}

\newcommand{\hipBtrend}{\ensuremath{13.6^{+0.3}_{-0.3}}\xspace}
\newcommand{\hipBcurv}{\ensuremath{\equiv 0}\xspace}
\newcommand{\hipBdmu}{\ensuremath{0.35^{+0.05}_{-0.05}}\xspace}

\newcommand{\hipBpplEthraid}{\ensuremath{$22\%$}\xspace}
\newcommand{\hipBpBDEthraid}{\ensuremath{$78\%$}\xspace}
\newcommand{\hipBpstarEthraid}{\ensuremath{$0\%$}\xspace}

\newcommand{\hipBpplOcto}{\ensuremath{$0\%$}\xspace}
\newcommand{\hipBpBDOcto}{\ensuremath{$100\%$}\xspace}
\newcommand{\hipBpstarOcto}{\ensuremath{$0\%$}\xspace}

\newcommand{\hipBOctoA}{\ensuremath{17.9^{+4.8}_{-2.7}}\xspace}
\newcommand{\hipBOctoM}{\ensuremath{45.2^{+10.5}_{-12.7}}\xspace}

\newcommand{\hipBEthraidA}{\ensuremath{12.6-49.9}\xspace}
\newcommand{\hipBEthraidM}{\ensuremath{3.8-64.2}\xspace}

\newcommand{\hipCtrend}{\ensuremath{4.3^{+0.8}_{-0.8}}\xspace}
\newcommand{\hipCcurv}{\ensuremath{-1.1^{+0.4}_{-0.4}}\xspace}
\newcommand{\hipCdmu}{\ensuremath{0.51^{+0.03}_{-0.03}}\xspace}

\newcommand{\hipCpplEthraid}{\ensuremath{$30\%$}\xspace}
\newcommand{\hipCpBDEthraid}{\ensuremath{$65\%$}\xspace}
\newcommand{\hipCpstarEthraid}{\ensuremath{$5\%$}\xspace}

\newcommand{\hipCpplOcto}{\ensuremath{$65\%$}\xspace}
\newcommand{\hipCpBDOcto}{\ensuremath{$35\%$}\xspace}
\newcommand{\hipCpstarOcto}{\ensuremath{$0\%$}\xspace}

\newcommand{\hipCOctoA}{\ensuremath{6.4^{+0.6}_{-0.3}}\xspace}
\newcommand{\hipCOctoM}{\ensuremath{9.5^{+5.4}_{-2.2}}\xspace}

\newcommand{\hipCEthraidA}{\ensuremath{6.9-58.0}\xspace}
\newcommand{\hipCEthraidM}{\ensuremath{8.8-120.7}\xspace}

\newcommand{\Msun}{\ensuremath{M_{\odot}}\xspace}
\newcommand{\Mjup}{\ensuremath{M_{\mathrm{Jup}}}\xspace}

\begin{document}



\title{Gaia Exoplanet Orbits, Demographics, and Evolution Survey (GEODES). I. \\ Characteristics of Three Long-Period Companions Accelerating their Host Stars}

\correspondingauthor{Judah Van Zandt}
\email{judahvz@ucsb.edu}

\author[0000-0002-4290-6826]{Judah Van Zandt}
\affiliation{Department of Physics, University of California, Santa Barbara, Santa Barbara, CA 93106, USA}

\author[0000-0003-2649-2288]{Brendan P. Bowler}
\affiliation{Department of Physics, University of California, Santa Barbara, Santa Barbara, CA 93106, USA}

\author[0000-0003-0967-2893]{Erik A. Petigura}
\affiliation{Department of Physics \& Astronomy, University of California Los Angeles, Los Angeles, CA 90095, USA}

\author[0000-0003-4557-414X]{Kyle Franson}
\altaffiliation{NHFP Sagan Fellow}
\affiliation{Department of Astronomy $\&$ Astrophysics, University of California, Santa Cruz, CA 95064, USA}

\author[0000-0002-6618-1137]{Jerry W. Xuan}
\altaffiliation{51 Pegasi b Fellow}
\affiliation{Department of Earth, Planetary, and Space Science, 595 Charles E. Young Dr E, University of California, Los Angeles, CA 90095, USA}

\author[0000-0003-4022-6234]{Marvin Morgan}
\affiliation{Department of Physics, University of California, Santa Barbara, Santa Barbara, CA 93106, USA}

\author[0000-0003-2646-3727]{Lauren I. Biddle}
\affiliation{Department of Astrophysical Sciences, Princeton University, Princeton, NJ 08540, USA}

\author[0000-0003-2102-3159]{Rocio Kiman}
\affiliation{Department of Physics, University of California, Santa Barbara, Santa Barbara, CA 93106, USA}

\author[0000-0002-2696-2406]{Jingwen Zhang}
\affiliation{Department of Physics, University of California, Santa Barbara, CA 93106, USA}

\author[0000-0002-0531-1073]{Howard Isaacson}
\affiliation{{Department of Astronomy, University of California Berkeley, Berkeley CA 94720, USA}}

\author[0000-0003-3504-5316]{Benjamin Fulton}
\affiliation{NASA Exoplanet Science Institute/Caltech-IPAC, MC 314-6, 1200 E. California Blvd., Pasadena, CA 91125, USA}

\author[0000-0001-8638-0320]{Andrew W. Howard}
\affiliation{Department of Astronomy, California Institute of Technology, Pasadena, CA 91125, USA}

\author[0000-0003-0800-0593]{Justin R. Crepp}
\affiliation{University of Notre Dame, Department of Physics and Astronomy, Notre Dame, IN, USA}

\author[0000-0001-5684-4593]{William Thompson}
\affiliation{National Research Council of Canada Herzberg, Victoria, BC, V9E 2E7, Canada}

\author[0000-0001-5684-4593]{Dori Blakely}
\affiliation{National Research Council of Canada Herzberg, Victoria, BC, V9E 2E7, Canada}
\affiliation{Department of Physics and Astronomy, University of Victoria, Victoria, BC V8P 5C2, Canada}

\begin{abstract}
The upcoming release of $Gaia$ DR4 will yield thousands of giant planet candidates, eventually enabling studies of giant planet eccentricities, masses, and occurrence rates across a broad range of stellar host masses, metallicities, and ages. However, some of these planet candidates are expected to be false positives, and even genuine detections will require additional observations to precisely determine their orbits and masses. We present here the first results of the $Gaia$ Exoplanet Orbits, Demographics, and Evolution Survey (GEODES), an observational campaign to identify the most promising planet candidate hosts for pre-DR4 vetting and post-DR4 validation and characterization. In this paper we showcase three systems from our broader sample exhibiting both tangential and radial accelerations, each representing a distinct outcome of our survey strategy. We combine $Hipparcos$, $Hipparcos$-$Gaia$, $Gaia$ DR2, and $Gaia$ DR3 absolute astrometry with adaptive optics (AO) imaging and precision RVs to constrain companion masses and orbits. HIP 18512, a nearby (15.3 pc) K4V dwarf, hosts a low-mass stellar companion at $10\farcs87 \pm 0\farcs07$ (166 AU) which produces significant RV and astrometric accelerations on its host star. The RV trend and astrometric acceleration of the nearby (24.2 pc) K4V star HIP 45839, together with an AO imaging non-detection, constrain the companion to $a$ = \hipBOctoA AU ($P$ = 70--127 years) and $M$ = \hipBOctoM \Mjup. In the case of HIP 81991 (43.8 pc, G5V), the astrometric and RV data indicate that the companion has a separation of \hipCOctoA AU ($P$ = 14.4--17.7 years) and a mass of \hipCOctoM \Mjup, and is more likely a planet ($\hipCpplOcto$) than a brown dwarf ($\hipCpBDOcto$).
\end{abstract}

\section{Introduction}
\label{sec:intro}

Mapping exoplanet demographics has proven to be one of the most effective tools for understanding their physical and orbital characteristics, their formation channels, and how they dynamically evolve over time. Thousands of transit detections from NASA's \textit{Kepler} Space Telescope \citep{Borucki2010} and Transiting Exoplanet Survey Satellite (\textit{TESS}; \citealt{Ricker2015}) enabled studies of small planet occurrence \citep{Petigura2018}, the Radius Gap \citep{Fulton2017}, the `Peas-in-a-Pod' pattern \citep{Millholland2017, Weiss2018}, the prevalence of hot Jupiters (e.g. \citealt{Yee2025}), and the eccentricities of short-period Jovians and sub-Jovians \citep{Dong2021, Fairnington2025, Gilbert2025}. Meanwhile, ground-based radial velocity (RV) surveys have yielded insights into giant planet occurrence rates as a function of mass \citealt{Grether2006, VanZandt2026}, separation \citep{Fulton2021}, eccentricity \citep{Morgan2025, Weldon2025, Gilbert2026, Blunt2026}, and host star metallicity \citep{Fischer2005, Johnson2010, Giacalone2026}. (See \citealt{WinnPetigura2024} for a review of recent transit and Doppler results.) At wider separations, directly imaged giant planets have been used to establish the eccentricities \citep{Bowler2020}, rotation rates \citep{Hsu2026}, abundance patterns \citep{Xuan2024}, occurrence rates \citep{Bowler2016, Nielsen2019, Vigan2021}, and obliquities of distant gas giants \citep{Snellen2014, Bryan2018, Poon2024}. 

The precision of statistical constraints placed on exoplanet sub-populations is directly linked to sample size. The steady growth of the census of confirmed exoplanets, which recently surpassed 6000,\footnote{https://exoplanetarchive.ipac.caltech.edu/, accessed 23 February 2026 (\citealt{Christiansen2025}).} together with the combination of multiple detection techniques, has enabled the exploration of more nuanced population characteristics, including the distribution of stellar obliquities (\citealt{Albrecht2022, Knudstrup2024, Rice2024}), true planetary mass measurements (see, e.g., \citealt{Chontos2022, An2025, Wallace2025, VanZandt2026}), planet-planet mutual inclinations \citep[e.g.,][]{Xuan2020, Zhang2025}, and the relationship between close-in small planets and long-period giants \citep{Bryan2025, VanZandt2025, Bonomo2025}.

The upcoming fourth data release from the European Space Agency's $Gaia$ mission \citep{Gaia2016}, $Gaia$ DR4, will precipitate an enormous leap in the number of cold Jupiters suitable for demographics \citep{Brown2025}. \cite{Lammers2026} has forecasted $\sim$7,500 new discoveries from DR4 and $\sim$120,000 from the future fifth data release, DR5. In principle, this will enable precise mapping of the cold Jupiter occurrence rate, eccentricity distribution, radial distribution, and stellar obliquity distribution---and how these vary with stellar host mass, metallicity, age, and multiplicity.


The transformative impact that $Gaia$ is poised to have on the exoplanet community is contingent on adequate preparation and observational follow-up. A number of false positive scenarios must first be excluded to confirm a $Gaia$ planet candidate. The most prominent origin of false-positive planet-like signals is expected to be from near equal-flux binaries with orbital periods from a few months to years. While the false positive fraction is not yet known, analyses of DR3 data and simulations of DR4 data predict a rate of $\sim$50-90\% \citep{Marcussen2023, Stefansson2025,Lammers2026} that will vary based on planet mass, semi-major axis, and stellar mass. Furthermore, only about one third of true detections will have orbits measured to 20\% or better by astrometry alone \citep{Lammers2026}, and even in confirmed cases, $Gaia$ astrometric solutions can depart from RV-based orbit fits (\citealt{Winn2022,Marcussen2023}). Finally, turning discoveries into demographics will require a detailed assessment of sample characteristics, sensitivities, biases in the orbit fits, and potential degeneracies in the interpretation of signals (e.g., \citealt{Yahalomi.2025}). Realizing the full potential of $Gaia$ DR4 thus depends on fastidious pre-release vetting for binary companions, as well as extensive follow-up RV and imaging efforts to confirm detected substellar companions and precisely measure their orbits and masses.
 
The $Gaia$ Exoplanet Orbits, Demographics, and Evolution Survey (GEODES) is an observational collaboration to identify, validate, and characterize the most promising potential hosts of planetary and substellar companions observed by $Gaia$. In preparation for DR4, we began a pilot survey of stars with astrometric variations suggestive of a low-mass companion. This initial sample was selected in part based on new orbit posteriors using additional astrometric constraints provided in the $Gaia$ DR2, DR3, and $Hipparcos$ (G23H; \citealt{Thompson2026}) catalogs, which have resulted in hundreds of new candidate substellar companions that are suitable for follow-up observations right now (Thompson et al. 2026, in prep.).


Alongside our overview of GEODES, in this work we present the analysis of three targets in our survey --- HIP 18512, HIP 45839, and HIP 81991 --- that highlight the potential of our program to identify promising targets after the release of $Gaia$ DR4. All of these stars exhibit astrometric signals consistent with a substellar companion, and each represents a possible outcome of our observation and analysis procedure: we recover the stellar companion to HIP 18512, use imaging to conclude that the companion to HIP 45839 is most likely substellar, and determine that the companion to HIP 81991 is either a giant planet or a brown dwarf using only RVs and astrometry.

The remainder of this study is organized as follows. In Section \ref{sec:sample_characteristics} we describe the procedure used to construct our sample, introduce three targets of interest---HIP 18512, HIP 45839, and HIP 81991--- showing accelerations consistent with a companion, and summarize their host star properties. We present archival observations for each of these targets in Section \ref{sec:observations}, and jointly analyze the mass and orbit constraints of their companions in Section \ref{sec:analysis}. We present results for each system in Section \ref{sec:results}, and conclude in Section \ref{sec:conclusion}.

\section{Pre-DR4 Sample Characteristics}
\label{sec:sample_characteristics}

$Gaia$ DR4 is scheduled to be released in late 2026. The most important aspects of this release as it relates to exoplanet detection will be the epoch astrometry, catalog of planet candidates, and systems with deviations from linear proper motion (acceleration terms corresponding to long orbital periods).\footnote{$Gaia$ DR4 content can be found at \url{https://www.cosmos.esa.int/web/gaia/dr4}} Leading up to DR4, several approaches have been developed to identify substellar companions using the existing DR2 and DR3 catalogs by combining these with $Hipparcos$ astrometry to directly image planets \citep{Brandt2018, Brandt2021a, Kervella2019, Kervella2022}, using the RUWE goodness-of-fit metric and the astrometric excess noise (AEN) parameter to tabulate thousands of candidate planets \citep{Kiefer2025}, and incorporating information about DR2-DR3 proper motion differences into predictions \citep{Penoyre2022, Feng2024, Ribas2025, Thompson2026, Blakely2026}.

Inevitably, some of these signals will be produced by false positives, many of which can be readily identified with high-resolution spectroscopy to search for double- and single-lined spectroscopic binaries, precision RVs to detect reflex motion from the companion, and high-resolution imaging to identify distant stellar and brown dwarf companions. In this section we describe new observations focused on a ``pre-DR4" sample of potential exoplanet hosts for follow-up, and examine three interesting targets that are representative of the systems in our survey. This sample of candidate host stars, together with our preparatory observational campaign, will be valuable once $Gaia$ DR4 is released, providing first epoch observations to combine with the new data to confirm or discard candidates.

\subsection{Sample selection}

The aim of our program is to assess the nature of the astrometric signals accessible now (pre-DR4), identify promising systems that are likely to emerge as planet candidates in DR4, and begin to design a larger effort that will ultimately yield reliable demographic information about cold Jupiters and brown dwarfs. We have therefore drawn from three primary (and evolving) sources for our broader sample: (1) bright stars ($G \lesssim$ 11~mag) with $Hipparcos$-$Gaia$ accelerations consistent with intermediate-separation giant planets (\citealt{Brandt2021a}), (2) bright stars ($G \lesssim$ 11~mag) with reliable planet candidates identified from the G23H catalog (\citealt{Thompson2026}; Thompson et al., in prep.), and (3) somewhat fainter stars ($G <$ 14~mag) identified with a $Gaia$ Renormalized Unit Weight Error (RUWE)-based metric and Astrometric Excess Noise (AEN) parameter following the procedure of the GaiaPMEX catalog (\citealt{Kiefer2025}), except applied to all stellar masses and not limited to $>$0.5~$M_{\odot}$. Our current target pool comprises 2270 stars, the two middle quartiles (25$^{\text{th}}$--75$^{\text{th}}$ percentiles) of which span masses of 0.9--1.4 $\Msun$ (G8 to F3), distances of 53--118 pc, and $G$ magnitudes of 7.2--9.1.

Our sample prior to the DR4 release focuses on breadth and includes subsamples with specific goals---for instance, high- and low-mass stars, metal-rich and metal-poor stars, stars with well-determined rotation periods, and both nearby and distant stars. Our observations are focused on binary vetting, host star characterization, and early epochs of precision RVs. High-resolution spectroscopy and RVs are primarily being acquired with the Levy spectrograph at Lick Observatory's Automated Planet Finder telescope \citep{Vogt2014}. We perform spectroscopic vetting using both APF and the Keck Planet Finder spectrograph (KPF; \citealt{Gibson2016}) at the W. M. Keck Observatory. Adaptive optics imaging is currently being carried out with Keck Observatory's NIRC2 imager and ShaneAO at Lick Observatory. Together these are being used to establish projected rotational velocities, abundances, stellar ages, and multiplicity status---both for close-in spectroscopic binaries and more distant visual binaries. Below we provide additional details and considerations that are being incorporated into our target selection.

In some cases, false positives may already be vetted with existing datasets, and true positives (previously known planets) provide a means of assessing the reliability of these approaches. The G23H planet candidates and GaiaPMEX-based strategy (Sources 2 and 3 above) are described in detail in \citet{Thompson2026} and \citet{Kiefer2025}. Here, we include additional details about the sequential filters that have gone into generating targets for Source 1 from the $Hipparcos$-$Gaia$ Catalog of Accelerations \citep[HGCA,][]{Brandt2021a}. For each star in the catalog, systems with significant HGCA accelerations are retained, and constraints are generated in companion mass and separation following the approach in \cite{fransonAstrometricAccelerationsDynamical_2023}. Stars are prioritized if the astrometric acceleration is likely to be caused by a brown dwarf or planet and is expected to be recovered in DR4, which is determined using the \texttt{htof} tool \citep{brandtHtofNewOpensource_2021} with an assumed epoch precision of $120 \, \mathrm{\mu as}$. \texttt{htof} directly queries the $Gaia$ Observing Schedule Tool (GOST; \citealt{FernandezHernandez2022}) to ensure proper modeling of the spacecraft's scanning law. Targets are further prioritized if they have a significant GaiaPMEX \citep{Kiefer2025} detection, or a significant DR2-DR3 proper motion difference (based on the calibration in \citealt{Thompson2026}) on top of the $Hipparcos$ to DR3 proper motion difference (HGCA). We prioritize stars that do not show signs of binarity from three $Gaia$ DR3 RV statistics: RV-renormalized goodness of fit ($\texttt{rv\_renormalised\_gof}$), the $p$-value for RV constancy ($\texttt{rv\_chisq\_pvalue}$) and the difference between the largest and the smallest RV measurement ($\texttt{rv\_amplitude\_robust}$). Finally, the Washington Double Star catalog (WDS; \citealt{Mason2001}) is queried and a search for common proper motion companions within 10" in $Gaia$ is carried out to identify visual binaries that may be causing the astrometric accelerations.

A large number of targets have previous RV and direct imaging observations, and many joint astrometric and RV orbit fits have been presented for known substellar companions (e.g., \citealt{Zhang2024, An2025, VanZandt2025}). Here, we highlight HIP 18512, HIP 45839, and HIP 81991: three systems in our sample that have significant RV trends with multi-year baselines, which we describe in Section~\ref{sec:observations}. In the rest of this section we summarize the general properties of these three host stars.

\label{sec:stellar_characterization}

\subsection{HIP 18512 Stellar Properties}
\label{subsec:stellar_data_hip18512}
HIP~18512 (HD~24916; $Gaia$~DR3~3256334497479041024) is a K4V, $V=8.0$ mag star at a distance of $15.27 \pm 0.01~\mathrm{pc}$ \citep{Queiroz2023}. HIP~18512 has a \textit{Gaia} Final Luminosity Age Mass Estimator (FLAME; \citealt{BailerJones2013, Creevey2023}) mass of $0.71 \pm 0.04~M_\odot$, a metallicity of [Fe/H]$ = 0.17^{+0.04}_{-0.02}$~dex \citep{GaiaCollaboration2022}, and a moderate projected rotational velocity of $v\sin i$=$2.5\pm 0.5$ km/s \citep{Perdelwitz2024}. $Gaia$ EDR3 reports a renormalized unit weight error (RUWE) of 1.06 for this target, favoring a single-star astrometric fit \citep{GaiaEDR32021}. \citet{BoroSaikia2018} and \citet{Isaacson2024} report $\log R'_{\mathrm{HK}}$ values between $-4.42$ and $-4.54$, placing this star in the chromospherically ``active'' category defined by \cite{Henry1996}. Using $Hipparcos$ kinematic data, \citet{Montes2001} and \citet{Madsen2002} classify HIP 18512 as a member of the Ursa Major association, suggesting a young age of $\sim$300~Myr (\citealt{SoderblomMayor1993}). We provide properties of the host star in Table \ref{table:stellar_params}.

HIP~18512 is part of a visual binary system with an M2.5 V companion at a projected separation of $\approx$11\arcsec\ (168~AU, $\Delta V$=3.6 mag; \citealt{Bessel1990}; \citealt{Poveda1994}; \citealt{Mason2001}; \citealt{Montes2018}). The companion is resolved in the $Gaia$ EDR3 catalog, which provides a parallax of $65.49 \pm 0.04$ mas, compared with $65.43 \pm 0.02$ mas for the primary. The proper motions are also comparable: the secondary has $\mu_{\alpha} \cos (\delta)=-209.41 \pm 0.04$ mas/yr and $\mu_{\delta}=-139.73 \pm 0.03$ mas/yr, versus $\mu_{\alpha} \cos (\delta)=-185.71 \pm 0.02$ mas/yr and $\mu_{\delta}=-142.91 \pm 0.02$ mas/yr for the primary. These values correspond to a relative angular velocity of $23.91\pm0.04$ mas/yr. Assuming a secondary mass of 0.552 $\Msun$ (see Section \ref{subsec:results_hip18512}), the relative angular velocity expected from a circular, face-on orbit is 35.24 mas/yr, meaning that the measured proper motions are consistent with a bound orbit.

\subsection{HIP 45839 Stellar Properties}\label{subsec:stellar_data_hip45839 }

The K4V star HIP 45839 (HD 80632; $Gaia$ DR3 5746426720312132608) is located at 24.15 $\pm$ 0.01 pc \citep{HardegreeUllman2023}. HIP 45839 has a mass of $0.72 \pm 0.06~M_\odot$ \citep{YeePetigura2017} and a metallicity of [Fe/H]$ = 0.11 \pm 0.01$~dex \citep{Huson2025}. Like HIP 18512, HIP 45839 has a moderate rotational velocity ($v\sin i$=$2.3 \pm 0.4$ km/s; \citealt{Perdelwitz2024}), well-suited for precision RV observations. The star's estimated age of $4.07^{+9.42}_{-2.84}~\mathrm{Gyr}$ \citep{YeePetigura2017} is essentially unconstrained. The host star's RUWE is 0.98 \citep{GaiaEDR32021}, well below the typical and even conservative limits of 1.4 \citep{Kiefer2025} and 1.2 \citep{Bryson2020} used to identify unresolved binaries. \citet{BoroSaikia2018} derived a $\log R'_{\mathrm{HK}}$ value of $-4.66$, consistent with the star being chromospherically active. We summarize HIP 45839's properties in Table \ref{table:stellar_params}.

\subsection{HIP 81991 Stellar Properties}\label{subsec:stellar_data_hip81991}

HIP~81991 (41 Her; HD~151090; $Gaia$ DR3~4435689739087756800), is a metal-poor ([Fe/H]=$-0.27\pm0.01$ dex; \citealt{Jonsson2020}) G8 dwarf with a mass of $1.09^{+0.07}_{-0.13}~\Msun$ \citep{Queiroz2020} located at a distance of $43.90 \pm 0.07~\mathrm{pc}$ \citep{Jonsson2020}. The star is inactive, with a $\log R'_{\mathrm{HK}}$ of $-$5.19 \citep{Isaacson2010}, exhibits a RUWE of 1.06 \citep{GaiaEDR32021}, and has a rotational velocity of approximately 4.36--5.37 km/s \citep{Abdurrouf2022, Jonsson2020}. There is a distant companion to HIP 81991 listed in the WDS ($\rho=163''$, $\Delta$V$\sim$4). The companion corresponds to $Gaia$ DR3 4435683451255623808 (HIP 81988, Ross 643), and has a proper motion of $\mu_{\alpha} \cos (\delta)=-215.62 \pm 0.02$ mas/yr and $\mu_{\delta}=-258.55 \pm 0.02$ mas/yr, consistent with that of the primary, $\mu_{\alpha} \cos (\delta)=-213.50 \pm 0.02$ mas/yr and $\mu_{\delta}=-257.50 \pm 0.02$ mas/yr. The parallax of the companion, $\varpi=22.76\pm0.02$ mas/yr, is also consistent with that of HIP 81991, $\varpi=22.78\pm0.02$ mas/yr. Assuming the companion is bound, its projected separation is $\sim$7150 AU. We provide system properties of HIP 81991 in Table \ref{table:stellar_params}.

\begin{table*}[t]
\centering
\caption{System Parameters}
\label{table:stellar_params}
\begin{tabular}{lcc}
\hline
\hline
\textbf{Parameter} & \textbf{Value} & \textbf{Source} \\
\hline
 & \textbf{HIP 18512} & \\
\hline
 & \textbf{General} & \\
Other Names & HD 24916, HR 8974, Gaia DR3 3256334497479041024 & \\
RA & 03:57:28.70 & \citet{GaiaEDR32021}\\
Dec &  $-$01:09:34.07 & \citet{GaiaEDR32021} \\
Spectral Type & K4V & \citet{Houk1999} \\
$G$ mag &  7.7 & \citet{GaiaEDR32021} \\
Mass (\Msun) & 0.71 $\pm$ 0.04\tablenotemark{a} & \citet{Creevey2023}\\
$T_{\text{eff}}$ (K) & $4640^{+100}_{-70}$ & \citet{GaiaDR2_RV2018}\\
$v\sin i$ (km/s) & $2.5 \pm 0.5$ & \citet{Perdelwitz2024}\\
\hline
 & \textbf{Astrometry} & \\
Parallax (mas) & 65.43 $\pm$ 0.02 & \citet{GaiaEDR32021} \\
Distance (pc) & 15.27 $\pm$ 0.01 & \citet{Queiroz2023} \\
$\mu_{\alpha}\cos(\delta)$ (mas/yr) & $-$185.71 $\pm$ 0.02 & \citet{GaiaEDR32021} \\
$\mu_{\delta}$ (mas/yr) & $-$142.91 $\pm$ 0.02 & \citet{GaiaEDR32021} \\
Radial Velocity (km/s) & 3.4 $\pm$ 0.1 & \citet{GaiaEDR32021} \\
RUWE & 1.06 & \citet{GaiaEDR32021} \\
\hline
 & \textbf{HIP 45839} & \\
\hline
 & \textbf{General} & \\
Other Names &  HD 80632, GJ 9296, Gaia DR3 5746426720312132608 & \\
RA & 09:20:44.32 & \citet{GaiaEDR32021}\\
Dec &  $-$05:45:14.30 & \citet{GaiaEDR32021} \\
Spectral Type & K4V & \citet{Houk1999} \\
$G$ mag &  8.7 & \citet{GaiaEDR32021} \\
Mass (\Msun) & 0.72 $\pm$ 0.06 & \citet{YeePetigura2017} \\
$T_{\text{eff}}$ (K) & $4550^{+130}_{-100}$ & \citet{GaiaDR2_RV2018}\\
$v\sin i$ (km/s) & $2.3 \pm 0.4$ & \citet{Perdelwitz2024}\\
\hline
 & \textbf{Astrometry} & \\
Parallax (mas) & 41.42 $\pm$ 0.02 & \citet{GaiaEDR32021} \\
Distance (pc) & 24.15 $\pm$ 0.01 & \citet{HardegreeUllman2023} \\
$\mu_{\alpha}\cos(\delta)$ (mas/yr) & $-$366.02 $\pm$ 0.02 & \citet{GaiaEDR32021} \\
$\mu_{\delta}$ (mas/yr) & $-$116.56 $\pm$ 0.02 & \citet{GaiaEDR32021} \\
Radial Velocity (km/s) & 37.0 $\pm$ 0.1 & \citet{GaiaEDR32021} \\
RUWE & 0.98 & \citet{GaiaEDR32021} \\
\hline
 & \textbf{HIP 81991} & \\
\hline
 & \textbf{General} & \\
Other Names &  HD 151090, 41 Her, Gaia DR3 4435689739087756800  & \\
RA & 16:44:59.99 & \citet{GaiaEDR32021}\\
Dec &  +06:05:17.36 & \citet{GaiaEDR32021} \\
Spectral Type & G8V & \citet{Bidelman1985} \\
$G$ mag &  6.3 & \citet{GaiaEDR32021} \\
Mass (\Msun) & $1.09^{+0.07}_{-0.13}$ & \citet{Queiroz2020} \\
$T_{\text{eff}}$ (K) & $5100^{+40}_{-70}$ & \citet{GaiaDR2_RV2018}\\
$v\sin i$ (km/s) & $5.37$ & \citet{Jonsson2020}\\
\hline
 & \textbf{Astrometry} & \\
Parallax (mas) & 22.78 $\pm$ 0.02 & \citet{GaiaEDR32021} \\
Distance (pc) & 43.90 $\pm$ 0.07 & \citet{Jonsson2020} \\
$\mu_{\alpha}\cos(\delta)$ (mas/yr) & $-$213.50 $\pm$ 0.02 & \citet{GaiaEDR32021} \\
$\mu_{\delta}$ (mas/yr) & $-$257.50 $\pm$ 0.02 & \citet{GaiaEDR32021} \\
Radial Velocity (km/s) & $-$6.6 $\pm$ 0.1 & \citet{GaiaEDR32021} \\
RUWE & 1.06 & \citet{GaiaEDR32021} \\
\hline
\hline
\end{tabular}
\tablenotetext{a}{\raggedright Mass calculated using the Final Luminosity Age Mass Estimator (FLAME; \citealt{BailerJones2013, Creevey2023}).}
\end{table*}

\section{Observations}
\label{sec:observations}

\subsection{HIP 18512}
\label{subsec:obs_hip18512}

\subsubsection{Radial Velocities}
\label{subsubsec:radial_velocities_hip18512}
We gathered literature RV measurements for HIP 18512 from the Data and Analysis Center for Exoplanets (DACE; \citealt{Buchschacher2019})\footnote{https://dace.unige.ch}. The RVs were collected at two facilities, Keck/HIRES \citep{Vogt1994} and La Silla/HARPS \citep{Mayor2003}. These instruments underwent upgrades in 2004 (see, e.g., \citealt{Butler2006}) and 2015 \citep{LoCurto2015}, respectively, resulting in four effective RV data sets for this system. The summary statistics for these time series are as follows: 13 pre-2004 HIRES RVs collected between October 1996 and September 2003 with median uncertainty 1.5 m/s, 17 post-2004 HIRES RVs collected between August 2004 and December 2018 with median uncertainty 1.1 m/s, 8 pre-2015 HARPS RVs collected in November 2014 with median uncertainty 0.7 m/s, and 16 post-2015 HARPS RVs collected between July 2015 and January 2017 with median uncertainty 0.7 m/s. Although the post-2015 HARPS data set originally contained 18 RVs, we removed two measurements with RV uncertainties of $\sim$20 m/s. The pre- and post-2004 HIRES RVs were published in \cite{Rosenthal2021} and the pre- and post-2015 HARPS RVs were published in \cite{Trifonov2020}. These data sets exhibit non-periodic RV variation on the order of 30 m/s over 23 years ($-$1.2 m/s/yr). We provide the measured RV trends in Table \ref{tab:trends}.


\begin{figure*}
    \centering
    \includegraphics[width=0.8\linewidth] {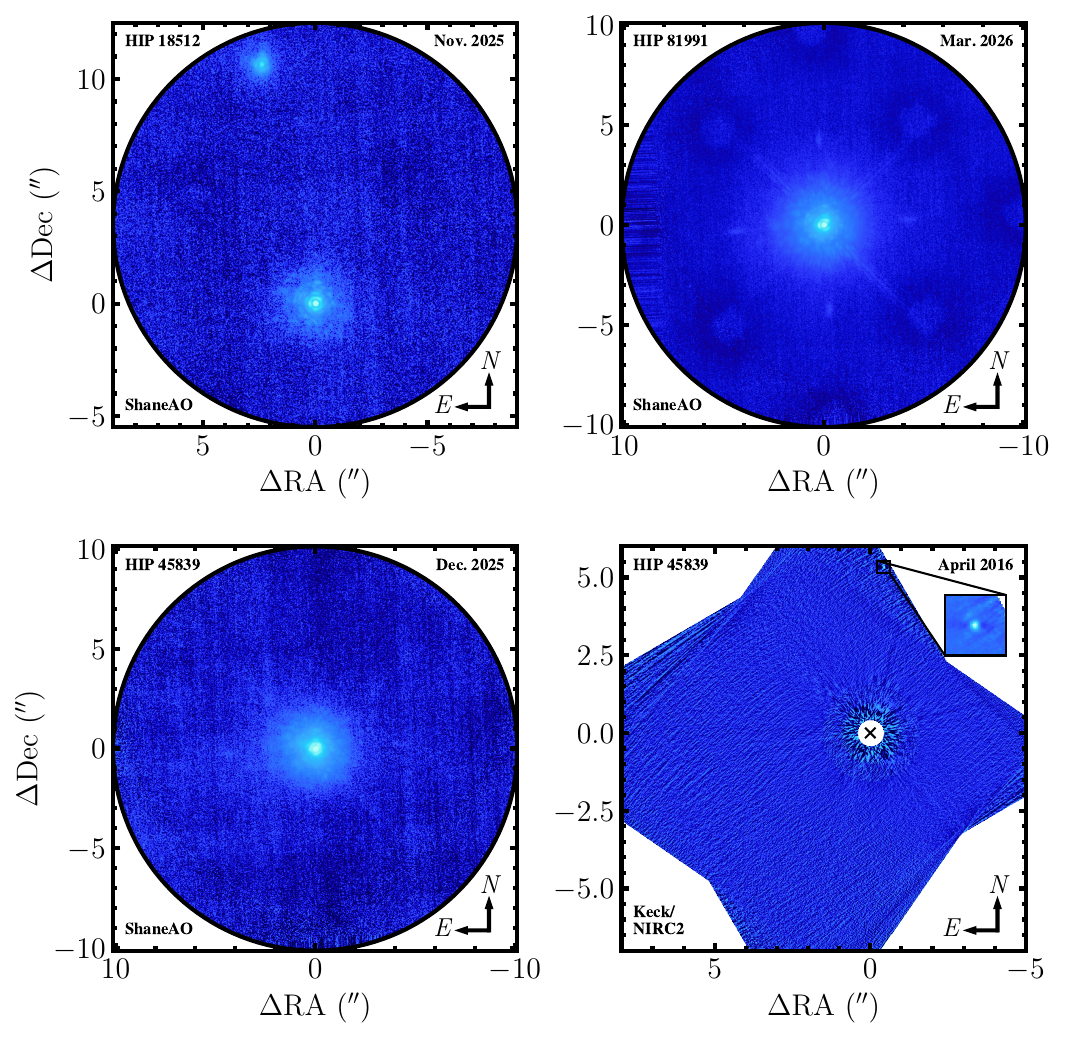}
    \caption{ShaneAO/ShARCS high-resolution imaging of HIP 18512 (left), ShaneAO/ShARCS imaging of HIP 45839 (center), and  Keck/NIRC2 high-contrast imaging of HIP 45839 (right). Each image is oriented such that north is up and and east is to the left. HIP 18512 shows a binary at a separation of 10\farcs9 and a position angle of $12.7^\circ$. The Keck/NIRC2 imaging of HIP 45839 shows a faint source at 5\farcs3 which is not consistent with the observed RV trend and astrometric acceleration. A zoomed-in $0\farcs4 \times 0\farcs4$ cutout of the source is shown in the inset panel. Images are displayed with an arcsinh stretch to capture a large dynamic range \citep{luptonPreparingRedGreen_2004}.\label{fig:imaging}}
\end{figure*}

\begin{figure}
    \centering
    \includegraphics[width=\linewidth] {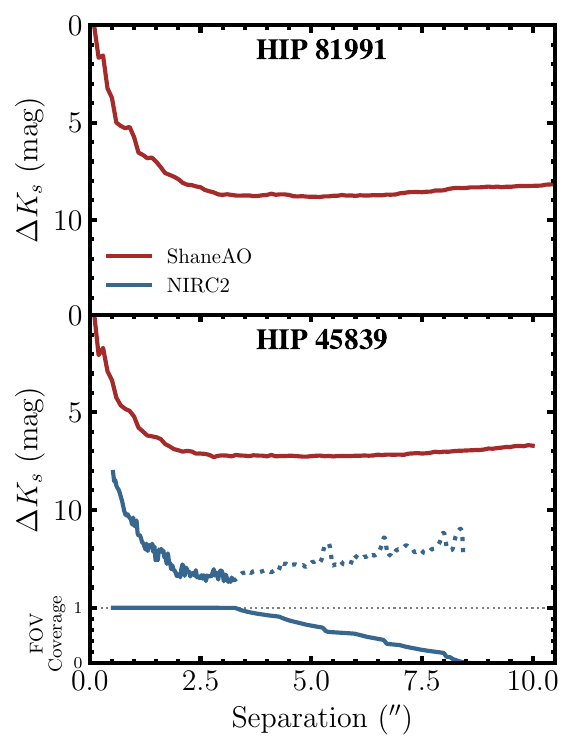}
    \caption{Contrast curves for HIP 81991 (top) and HIP 45839 (bottom) showing $5\sigma$ sensitivity limits from ShaneAO (red) and archival Keck/NIRC2 (blue) imaging. At a given angular separation, a companion with lower contrast, i.e. smaller $\Delta K_s$, than the value indicated by either curve would be detectable in the corresponding imaging data at $>5\sigma$ significance. Regions of partial field of view (FOV) coverage for the Keck/NIRC2 data are shown as the dotted line, with the FOV coverage shown in the bottom sub-panel. On average, regions at higher angular separation were observed in fewer frames of the ADI sequence, leading to a lower signal-to-noise ratio and thus lower derived contrast.}
    \label{fig:contrast_curves}
\end{figure}

\subsubsection{Imaging}
\label{subsubsec:imaging_hip18512}
We obtained natural guide star Adaptive Optics (AO) imaging of HIP 18512 with the ShARCS camera and ShaneAO \citep{srinathSwimmingSharcsComparison_2014} on the 3-m Shane Telescope at Lick Observatory on UT 2025 November 2. Our observations consist of five $t_\mathrm{int} = 4.4\, \mathrm{s}$ exposures in $K_s$ within a five-point box dither pattern. To avoid saturation, we used a two-magnitude neutral density filter (ND2). After AO correction, the FWHM of the host-star is $160 \, \mathrm{mas}$. Our data are reduced using a custom pipeline detailed in Franson et al. (in prep.). Briefly, raw frames are flat-fielded and dark subtracted. Bad pixels are flagged through a custom bad pixel mask (P. Lynam, priv. communication) and the \texttt{ccdmask} routine in \texttt{ccdproc} \citep{craigAstropyCcdprocV130post1_2017}. We then subtract the sky background, here taken as the median of the dithered frames.

These data show the known wide-orbiting companion at ${\sim}11^{\prime\prime}$. The source falls off the field of view in two dithers, so we exclude these frames from our analysis. To measure the relative astrometry and photometry of the companion, we fit 2D elliptical Gaussians to the Point Spread Functions (PSFs) of the host star and companion in each raw frame. To calculate separation and position angle, we take the mean of the centroid positions across all included frames. We use the standard deviation of these measurements as an estimate of uncertainty. We then transform the measurement from relative pixels to separation in angular units $\rho$ and north-aligned position angle $\theta$ through the ShARCS plate scale $s = 32.721 \pm 0.033 \, \mathrm{mas/px}$ and north alignment angle $\theta_\mathrm{north} = 1.87 \pm 0.13^\circ$ determined in Franson et al. (in prep.). Relative photometry of the companion is measured by performing aperture photometry within a 1 FWHM-radius circular aperture on the companion and host-star in each frame. The mean and standard deviation is then taken for the companion contrast $\Delta K_s$ and uncertainty. We measure $\rho = 10\farcs87 \pm 0\farcs07$, $\theta = 12.73 \pm 0.31^\circ$, and $\Delta K_s = 2.07 \pm 0.30 \, \mathrm{mag}$. Median-combined imaging of the binary is shown in Figure \ref{fig:imaging}.

We assess our sensitivity to companions as a function of angular separation by calculating the standard deviation of the pixels within concentric annuli in the final image. At each angular separation, the $5\sigma$ contrast threshold is defined as five times the standard deviation in that annulus, normalized to the sky-subtracted peak flux of the primary star. This procedure produces the contrast curve given in Table \ref{tab:contrast_curves}.

\begin{deluxetable}{cccc}
\tabletypesize{\footnotesize}
\tablecaption{Contrast Curves for All Targets}
\label{tab:contrast_curves}

\tablehead{
    \colhead{Target} &
    \colhead{Instrument} &
    \colhead{Angular Separation (")} &
    \colhead{Contrast ($\Delta K_s$)}
}

\startdata
HIP 18512 &   ShARCS &     0.10 &     0.068 \\
HIP 18512 &   ShARCS &     0.20 &     2.051 \\
HIP 18512 &   ShARCS &     0.30 &     1.915 \\
HIP 18512 &   ShARCS &     0.40 &     3.610 \\
HIP 18512 &   ShARCS &     0.50 &     4.322 \\
\hline
HIP 45839 &   ShARCS &     0.10 &    -0.062 \\
HIP 45839 &   ShARCS &     0.20 &     2.057 \\
HIP 45839 &   ShARCS &     0.30 &     1.684 \\
HIP 45839 &   ShARCS &     0.40 &     2.880 \\
HIP 45839 &   ShARCS &     0.50 &     3.335 \\
\hline
HIP 45839 &    NIRC2 &    0.53 &     8.540 \\
HIP 45839 &    NIRC2 &    0.55 &     8.471 \\
HIP 45839 &    NIRC2 &    0.58 &     8.771 \\
HIP 45839 &    NIRC2 &    0.60 &     8.887 \\
HIP 45839 &    NIRC2 &    0.63 &     8.970 \\
\hline
HIP 81991 &   ShARCS &     0.10 &     0.034 \\
HIP 81991 &   ShARCS &     0.20 &     1.669 \\
HIP 81991 &   ShARCS &     0.30 &     1.546 \\
HIP 81991 &   ShARCS &     0.40 &     3.236 \\
HIP 81991 &   ShARCS &     0.50 &     3.723 \\
\enddata
\tablecomments{We show the form of each target's contrast curve in this table, and provide the full contrast curves in the online version of this article.}

\end{deluxetable}

\subsection{HIP 45839}
\label{subsec:obs_hip45839}

\subsubsection{Radial Velocities}
\label{subsubsec:radial_velocities_hip45839}

We assembled literature RVs of HIP 45839 from HARPS \citep{Trifonov2020} and HIRES \citep{Butler2017}. The 23 pre-2015 HARPS RVs were collected between February 2004 and March 2015 and have a median uncertainty of 2.0 m/s, while the 8 post-2015 HARPS RVs, collected between February 2016 and March 2021, have a median uncertainty of 2.2 m/s. The five HIRES RVs span from December 2008 to February 2013 and have a median RV uncertainty of 1.0 m/s. These RVs show a strong acceleration of 13.6 m/s/yr.

We also observed HIP 45839 with the Automated Planet Finder (APF; \citealt{Vogt2014}). APF consists of a 2.4-m robotic telescope mounted with the Levy echelle spectrometer. Its RV reduction pipeline was adapted from the one used for HIRES and uses the same forward modeling approach \citep{Howard2010, Rosenthal2021}. We observe HIP 45839 using two consecutive exposures per observation, each reaching a signal-to-noise ratio (SNR) per pixel of 60 on blaze at 560 nm, for a total SNR of $\sim$85, equivalent to an RV precision of 3 m/s. While our aim is to extend the RV baseline of this target with APF, we have so far obtained only a handful of RVs, collected between February--March 2026. We will present these measurements in a future study after establishing a sufficient baseline to measure long-term RV reflex motion. We used one RV, collected on UT 2026 February 7, to evaluate the possibility that HIP 45839 is a double-lined spectroscopic binary (SB2). We cross-correlated the observation with the binary mask used in the ESPRESSO pipeline \citep{Pepe2021} and found only one peak in the cross-correlation function (CCF), indicating that there is not a second stellar companion at close separation.

\subsubsection{Imaging}
\label{subsubsec:imaging_hip45839}
We obtained ShaneAO/ShARCS imaging of HIP 45839 on UT 2025 December 4. Our data consists of five $1.5 \, \mathrm{s}$ $K_s$ exposures taken in a five-point box dither pattern. The FWHM of the host-star after AO correction is $230 \, \mathrm{mas}$. Here, we apply the same basic reduction steps as in Section \ref{subsubsec:imaging_hip18512}. The co-registered and median-combined reduced image is shown in Figure \ref{fig:imaging}, and the $5\sigma$ contrast curve is given in Figure \ref{fig:contrast_curves} and Table \ref{tab:contrast_curves}.

HIP 45839 was also targeted with the NIRC2 camera at W.M. Keck Observatory on UT 2016 April 21 as part of the TRENDS survey (PI: J. Crepp; \citealt{creppTrendsHighcontrastImaging_2012, creppTrendsHighcontrastImaging_2013, creppTrendsHighcontrastImaging_2013a, creppTrendsHighcontrastImaging_2014, creppTrendsHighcontrastImaging_2016, creppTrendsHighcontrastImaging_2018, montetTrendsHighcontrastImaging_2014, gonzalesTrendsHighcontrastImaging_2020}). The data were taken with NGS Adaptive Optics \citep{wizinowichAstronomicalScienceAdaptive_2013} and consist of 107 frames with the star behind the $600\, \mathrm{mas}$-diameter Lyot coronagraphic mask. The data were taken in pupil-tracking mode, which facilitates Angular Differential Imaging (ADI; \citealt{liuSubstructureCircumstellarDisk_2004,maroisAngularDifferentialImaging_2006}). Each frame is composed of seven $6\, \mathrm{s}$ coadds for a total on-source integration time of 74.9 min. A total of $25.1^\circ$ of field rotation was accrued across the sequence. Unsaturated images of the host star were periodically obtained, which we use for flux calibration. We derive a contrast curve for our NIRC2 imaging according to the same procedure as our ShaneAO curve (see Figure \ref{fig:contrast_curves} and Table \ref{tab:contrast_curves}). The variable field of view coverage in the NIRC2 images results in a lower achieved contrast at wide angular separations.

These data are reduced using a custom pipeline. Basic calibration steps are first applied: flat fielding, subtracting darks, identifying and masking cosmic rays using the \texttt{LACOSMIC} algorithm \citep{vandokkumCosmicRayRejection_2001}, and applying the NIRC2 distortion solution \citep{Service2016} using \texttt{rain}.\footnote{\href{https://github.com/jsnguyen/rain}{https://github.com/jsnguyen/rain}} We then co-register the frames by fitting a centroid to the host-star signal behind the partially-transparent mask. PSF subtraction is carried out through the implementation of the Karhunen-Lo\'eve Image Processing (KLIP; \citealt{soummerDetectionCharacterizationExoplanets_2012}) algorithm in \texttt{pyKLIP} \citep{wangPyklipPsfSubtraction_2015} with 3 annuli, 4 subsections, a movement parameter of 1 pixel, and 25 KL modes.

We detect a faint source near the edge of the field of view and apply the KLIP framework to extract its astrometry and photometry. We sample the separation ($\rho$), position angle ($\theta$), and contrast ($\Delta K_s$) parameter space using \texttt{emcee} \citep{foreman-mackeyEmceeMcmcHammer_2013} with 100 walkers and 2000 steps per walker, discarding the first 20\% of each chain as burn-in. Following \citet{fransonDynamicalMassYoung_2023}, uncertainties are determined by combining the measurement uncertainty from the posterior, the uncertainty on the \citet{Service2016} distortion solution of $\sigma_d = 1 \, \mathrm{mas}$, and the uncertainty on the plate scale and north-alignment angle. We also set a noise floor of $\pm5\,\mathrm{mas}$ in separation and $\pm 0.06^\circ$ in position angle to account for additional systematic uncertainty at the level of $4{-}5 \, \mathrm{mas}$ on the host-star position behind the $600 \, \mathrm{mas}$-diameter Lyot mask \citep[e.g.,][]{konopackyAstrometricMonitoringHr_2016,bowlerOrbitDynamicalMass_2018}. This produces the following astrometry and photometry for the source: $\rho = 5\farcs349 \pm 0\farcs005$, $\theta = -4.57^\circ \pm 0.06^\circ$, and $\Delta K_s = 6.23 \pm 0.07 \, \mathrm{mag}$. For the host-star's $K$-band magnitude in 2MASS of $6.268 \pm 0.029 \, \mathrm{mag}$, this corresponds to an apparent magnitude $K_s = 12.50 \pm 0.08 \, \mathrm{mag}$ and absolute magnitude $M_{K_s} = 10.58 \pm 0.08$ if the source shares the parallax of HIP 45839.

We measure a $5\sigma$ contrast curve by determining the raw contrast level in annuli out to the edge of the detector, applying the \citet{mawetFundamentalLimitationsHigh_2014} correction for small sample statistics at close separations. The source is masked in the contrast curve calculation. We then calibrate the algorithmic throughput through injection--recovery across five azimuthal angles. Here, the contrast curve calculation is run to the edge of the reduced image including areas with partial field of view (FOV) coverage. The resulting contrast curve is shown in Figure \ref{fig:contrast_curves}.

\subsection{HIP 81991}
\label{subsec:obs_hip81991}

\subsubsection{Radial Velocities}
\label{subsubsec:radial_velocities_hip81991}

The only archival RV measurements of HIP 81991 were obtained using the Hamilton spectrograph at Lick Observatory \citep{Vogt1987}. \cite{Fischer2014} presented 71 RVs of this star, acquired between July 2001 and September 2008, with a median RV uncertainty of 4.0 m/s. These measurements show long-term RV variation over seven years, with a linear component of \hipCtrend m/s/yr and a quadratic component of \hipCcurv m/s/yr$^2$ (see Table \ref{tab:trends}). As in the case of HIP 45839, we obtained an APF spectrum of HIP 81991 on UT 2026 March 1, reaching an SNR of 110 in 204 seconds. Using a binary mask, we found that the CCF is single-peaked, ruling out an SB2 scenario.

\subsubsection{Imaging}
\label{subsubsec:imaging_hip81991}
We obtained ShaneAO/ShARCS imaging of HIP 81991 on UT 2026 March 8. Our data comprise 47 single-coadd $K_s$ frames with $t_\mathrm{int} = 8.7 \, \mathrm{s}$ for a total on-source integration time of $6.8 \, \mathrm{min}$. The sequence is taken in a five-point box dither pattern. These data are reduced using the same pipeline described in Section \ref{subsubsec:imaging_hip18512}. The co-registered, median-combined reduced image is shown in Figure \ref{fig:imaging}. We present the $5\sigma$ contrast curve in Figure 
\ref{fig:contrast_curves} and in Table \ref{tab:contrast_curves}.

\subsection{Absolute Astrometry}
\label{subsec:absolute_astrometry}

The \textit{Hipparcos}-\textit{Gaia} Catalog of Accelerations (HGCA; \citealt{Brandt2021a}) placed the \textit{Hipparcos} \cite{Hipparcos1997} and \textit{Gaia} EDR3 \cite{GaiaEDR32021} catalogs in the same reference frame, facilitating comparison between stellar astrometric motion measurements over the $\sim$25-year baseline between these two missions. The HGCA reported significant astrometric proper motion anomalies (PMa) for HIP 18512 ($8\sigma$), HIP 45839 ($7\sigma$), and HIP 81991 ($17\sigma$). In the case of HIP 18512, the acceleration is consistent with having been caused by the known stellar companion (see Figure \ref{fig:ethraid_octo_HIP18512}). We list each star's PMa in Table \ref{tab:trends}.

\section{Analysis}
\label{sec:analysis}

\subsection{Radial Velocity Fits}
\label{subsec:rv_fits}

We fit the RVs for each of our three systems with \texttt{RVSearch} \citep{Rosenthal2021}. \texttt{RVSearch} identifies the most significant periodic signal in an RV time series, fits it with a Keplerian orbit model, subtracts the model from the data, and performs a new search on the residuals. The algorithm continues until no more significant signals are identified in the data. We also allowed for the inclusion of non-periodic linear and quadratic terms, which model long-period variation induced by distant companions. We found significant linear trends in all three systems, and list their values in Table \ref{tab:trends}. For HIP 81991, we fit a marginally significant ($2.8\sigma$) quadratic curvature term along with the trend. For other two systems, the quadratic term did not improve the model significantly, and was therefore fixed to zero.

\subsection{Joint RV/Astrometry Orbital Fit}
\label{subsec:orbit_fit}

We have described the RV, astrometric, and imaging datasets we assembled for each candidate. In this section, we aim to apply a homogeneous analysis of these three datasets for each candidate. All candidates have orbital periods exceeding the RV baselines. Fitting partial orbits requires care since the constraints are broad and prior dependent. We elected to use two codes with different modeling approaches as a consistency check.

\begin{figure*}[t]
    \centering
    
    \begin{minipage}{0.75\linewidth}
        \centering
        \includegraphics[width=\linewidth]{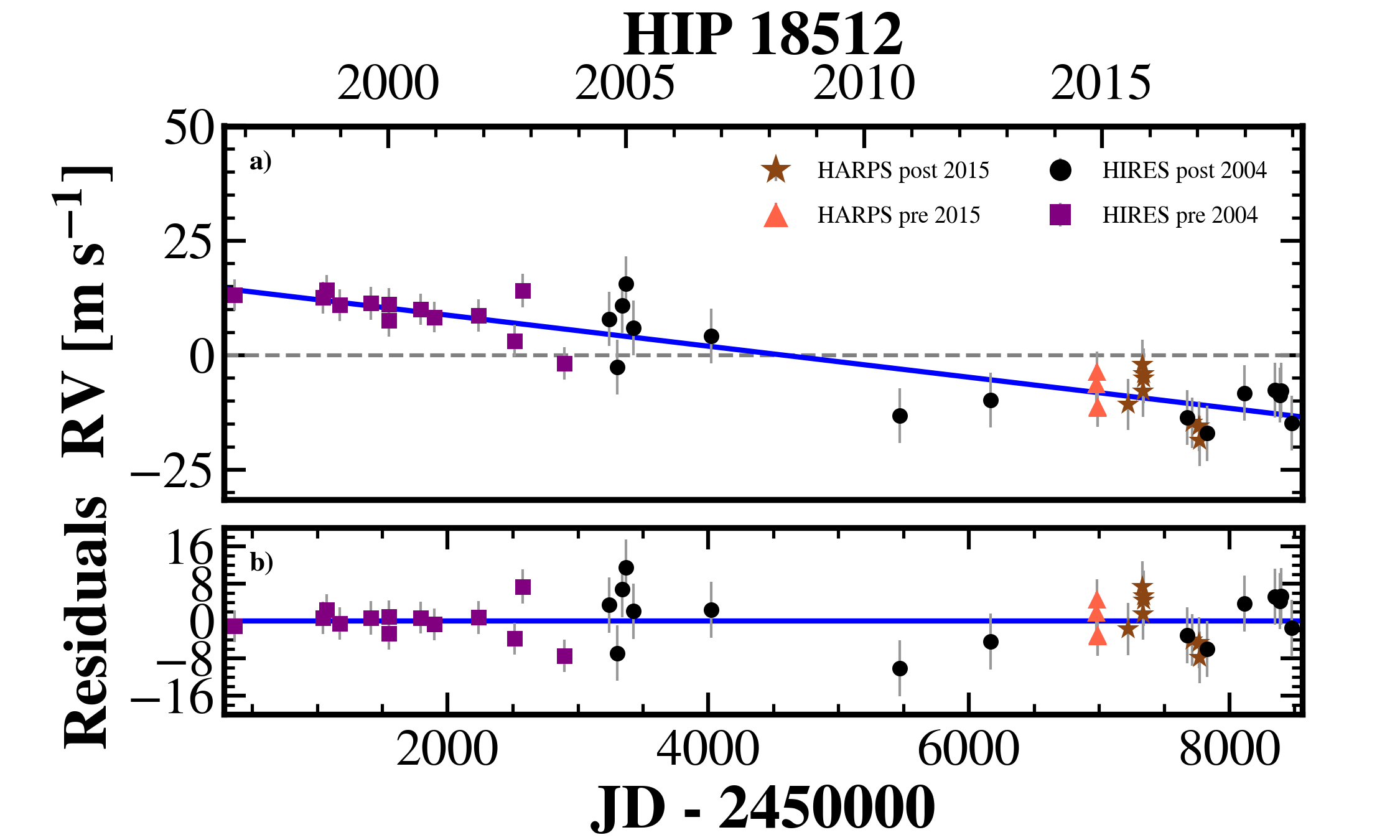}
    \end{minipage}
    \begin{minipage}{0.95\linewidth}
        \centering
        \includegraphics[width=\linewidth]{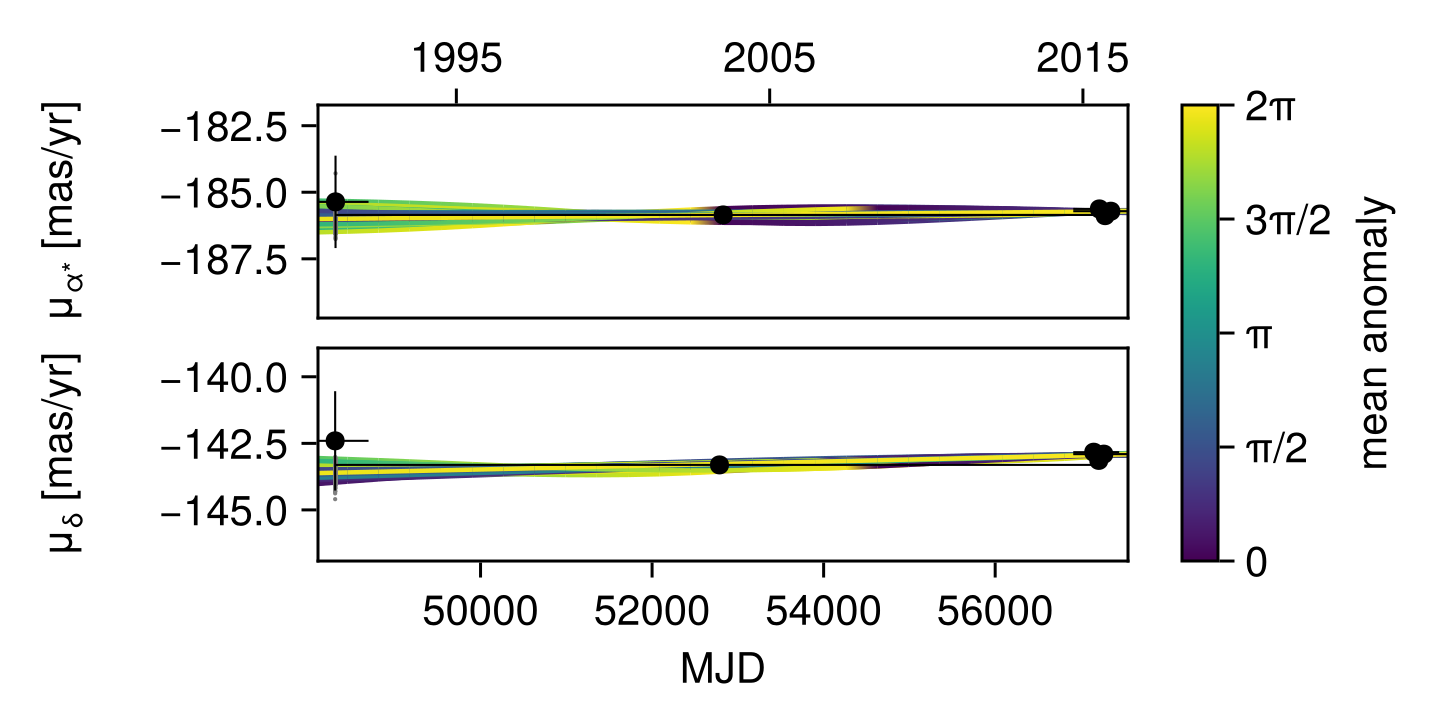}
    \end{minipage}
    
    \caption{\textbf{Top:} RV time series of HIP 18512. In Panel a), colored points and gray lines show the RV measurements and errors, respectively. The blue line shows the maximum likelihood model, with the RV residuals to this model plotted in Panel b). \textbf{Bottom:} Proper motion time-series from the joint astrometry+RV fit using \texttt{octofitter}. The colored curves are random posterior draws of the model. Black points show catalog measurements from the G23H catalog, including proper motions from $Hipparcos$, $Gaia$ DR2, and $Gaia$ DR3, as well as scaled positional differences between $Hipparcos$–$Gaia$ and DR2–DR3. Horizontal error bars indicate the time baselines over which each proper motion is measured.
    }
    \label{fig:rv_astro_fits_HIP18512}
\end{figure*}

\begin{figure*}[t]
    \centering
    \begin{minipage}[b]{0.462\linewidth}
        \centering
        \includegraphics[width=\linewidth]{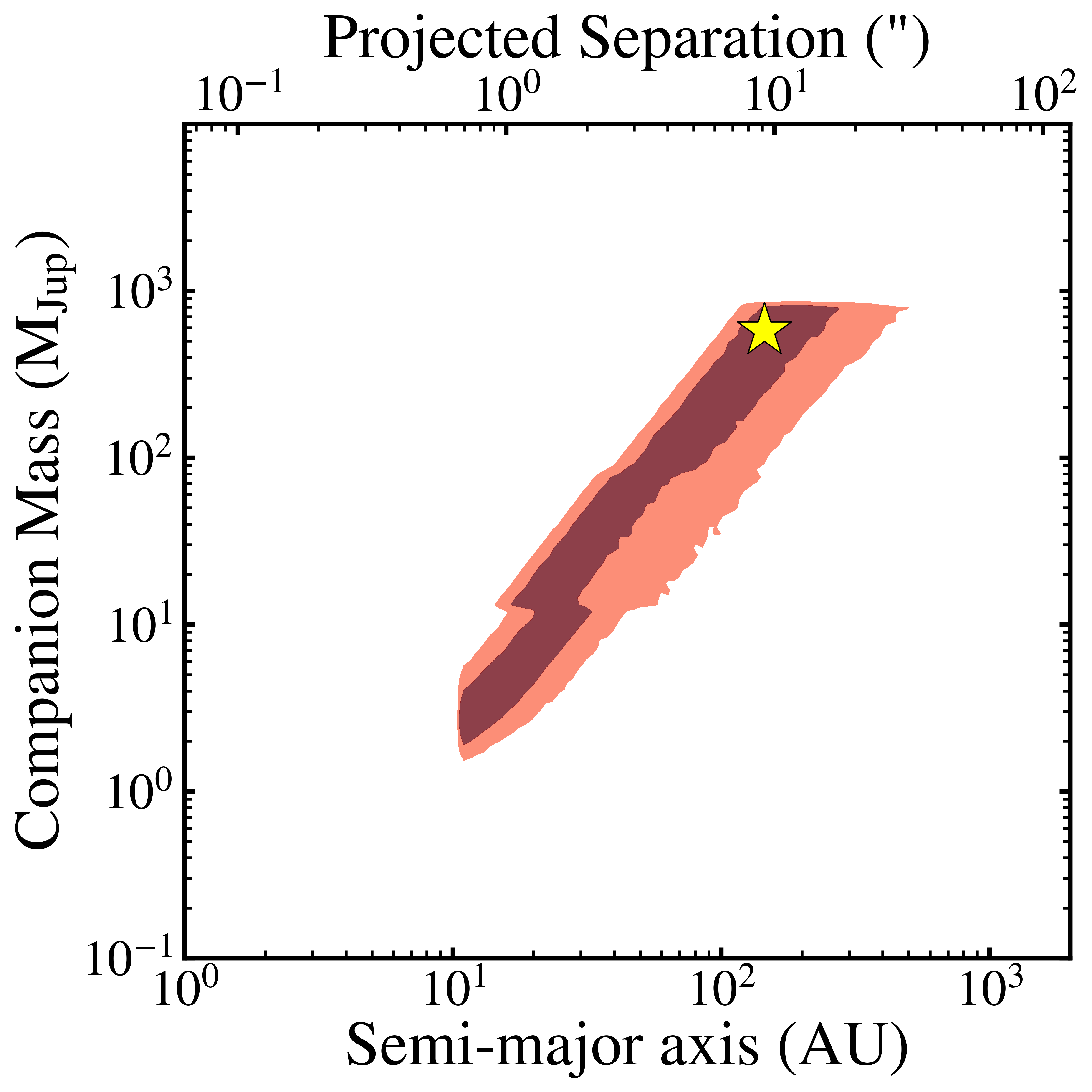}
    \end{minipage}
        \hfill
    \begin{minipage}[b]{0.52\linewidth}
        \centering
        \includegraphics[width=\linewidth]{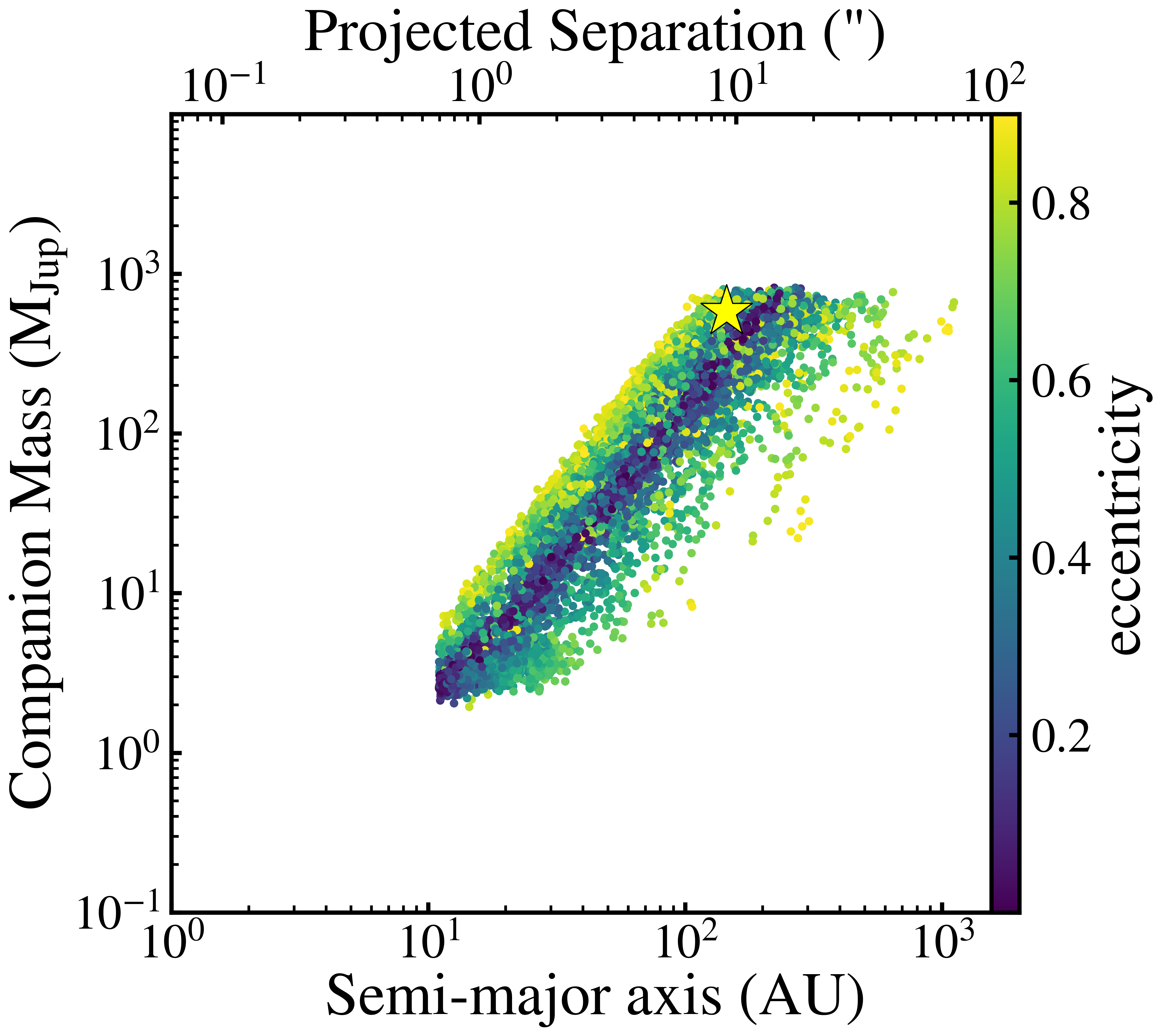}
    \end{minipage}
    
    \caption{\textbf{Left:} Our partial orbit fits to our RVs and absolute astrometry for HIP 18512. Red contours show constraints derived from \texttt{ethraid}, with the dark region covering 68\% of the posterior mass, and the light region covering 95\%. The gold star marks the mass and separation of the known M2.5 V stellar companion ($V=11.6$) derived by \cite{An2025}. We detected the companion in the AO images and find that its parameters are consistent with both the RVs and astrometry. \textbf{Right:} Posterior draws from our \texttt{octofitter} analysis, which all the data used by \texttt{ethraid} as well as $Gaia$ DR2-DR3 proper motions, $Gaia$ astrometric excess noise (AEN), and $Hipparcos$ intermediate astrometric data (IAD). Marker colors indicate orbital eccentricity.
    }
    \label{fig:ethraid_octo_HIP18512}
\end{figure*}

\begin{figure*}[t]
    \centering
    \begin{minipage}[b]{0.75\linewidth}
        \centering
        \includegraphics[width=\linewidth]{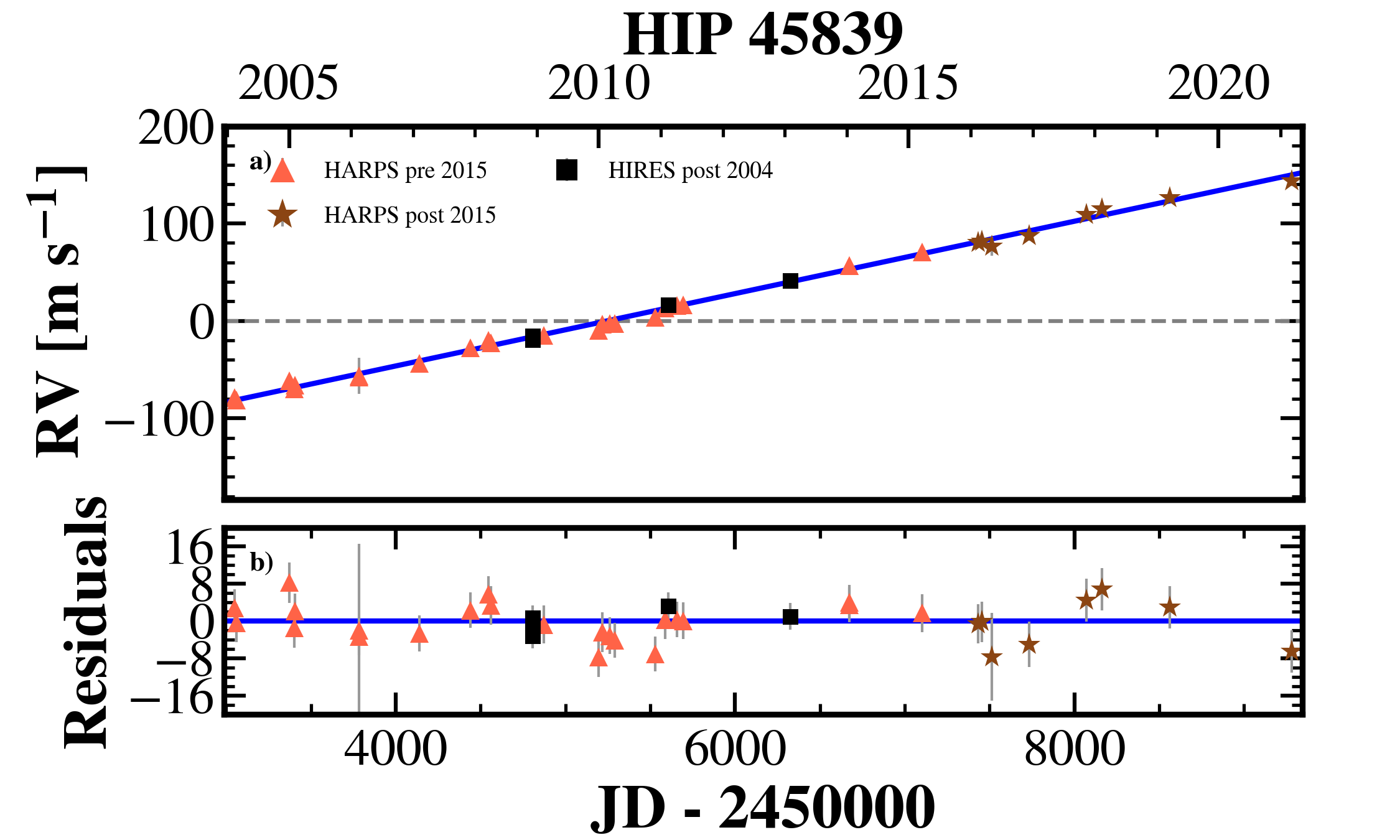}
    \end{minipage}
    \begin{minipage}[b]{0.95\linewidth}
        \centering
        \includegraphics[width=\linewidth]{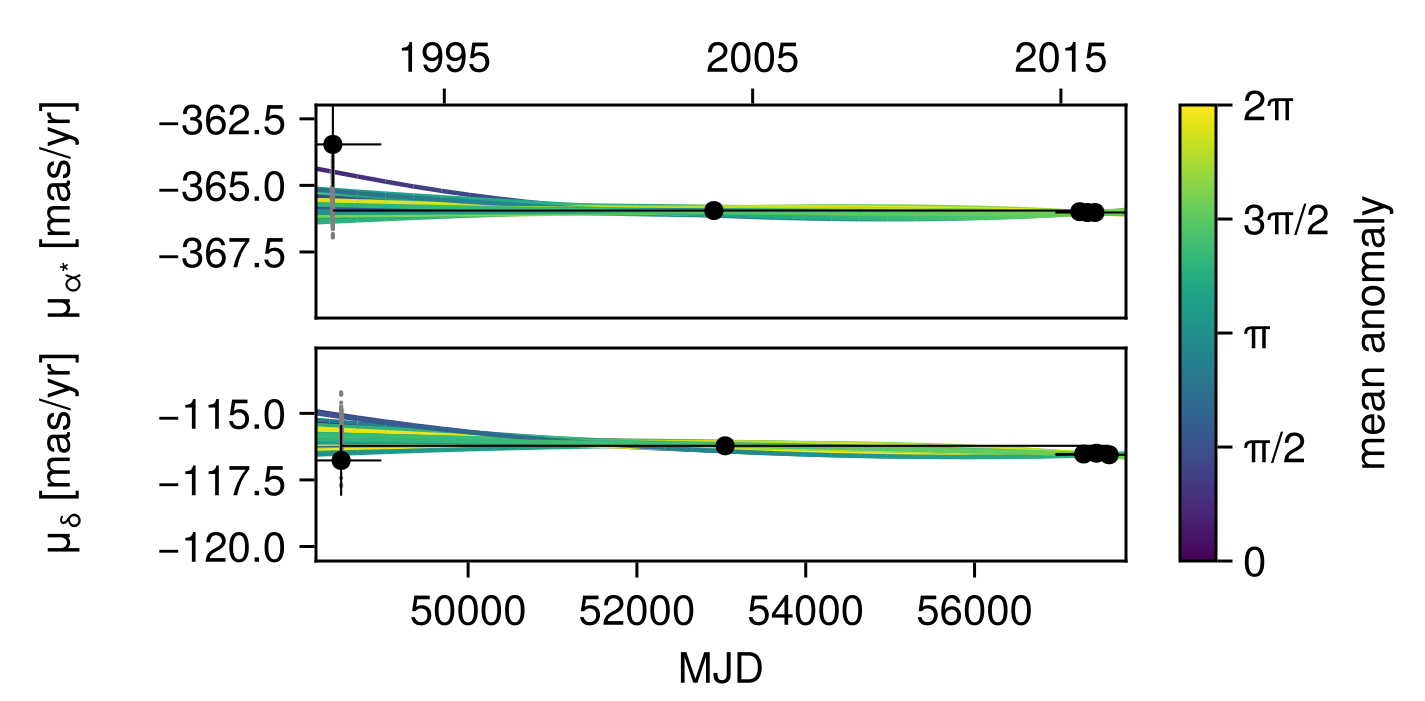}
    \end{minipage}
    
    \caption{Same as Figure \ref{fig:rv_astro_fits_HIP18512} for HIP 45839.}
    \label{fig:rv_astro_fits_HIP45839}
\end{figure*}

\begin{figure*}[t]
    \centering
    \begin{minipage}[b]{0.462\linewidth}
        \centering
        \includegraphics[width=\linewidth]{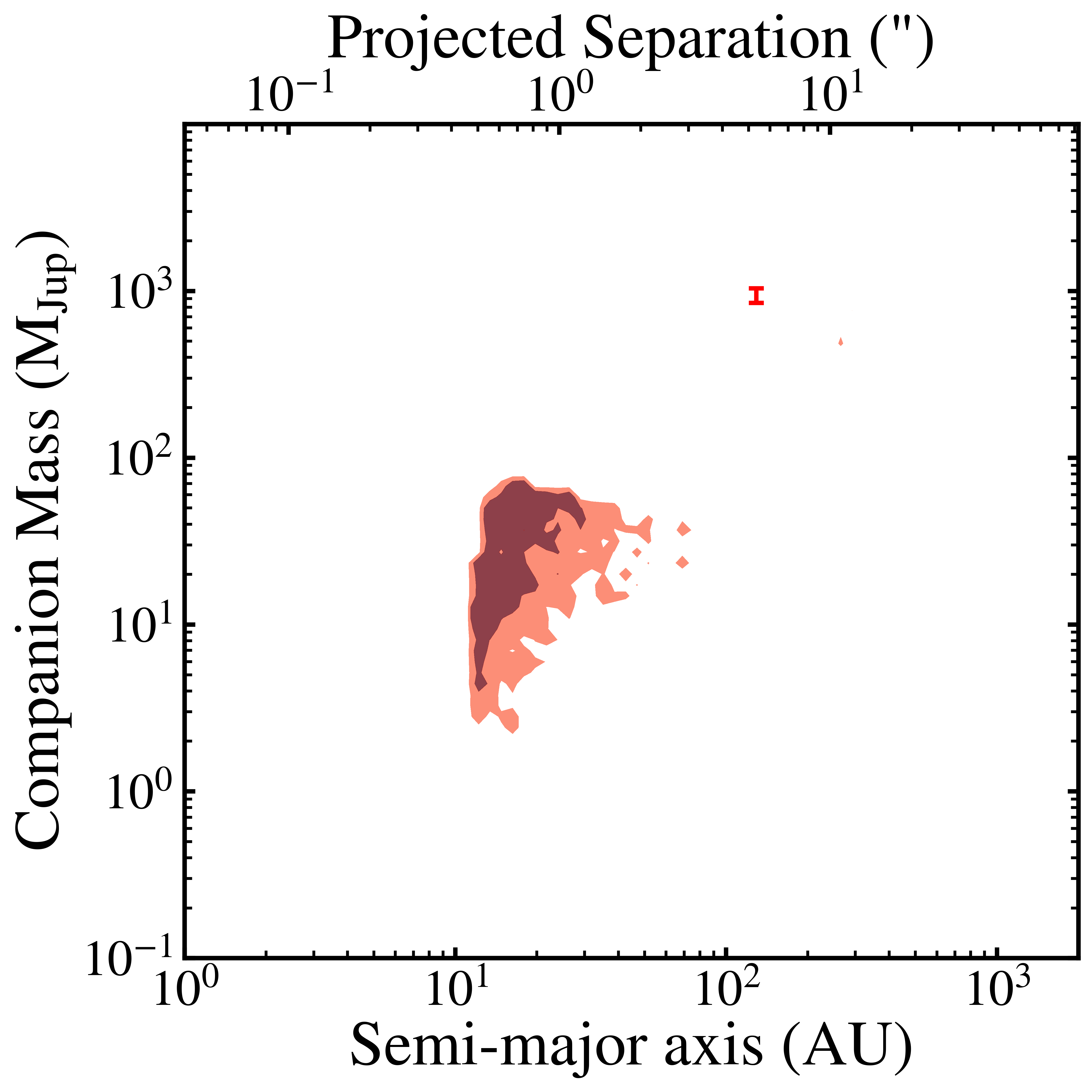}
    \end{minipage}
    \hfill
    \begin{minipage}[b]{0.52\linewidth}
        \centering
        \includegraphics[width=\linewidth]{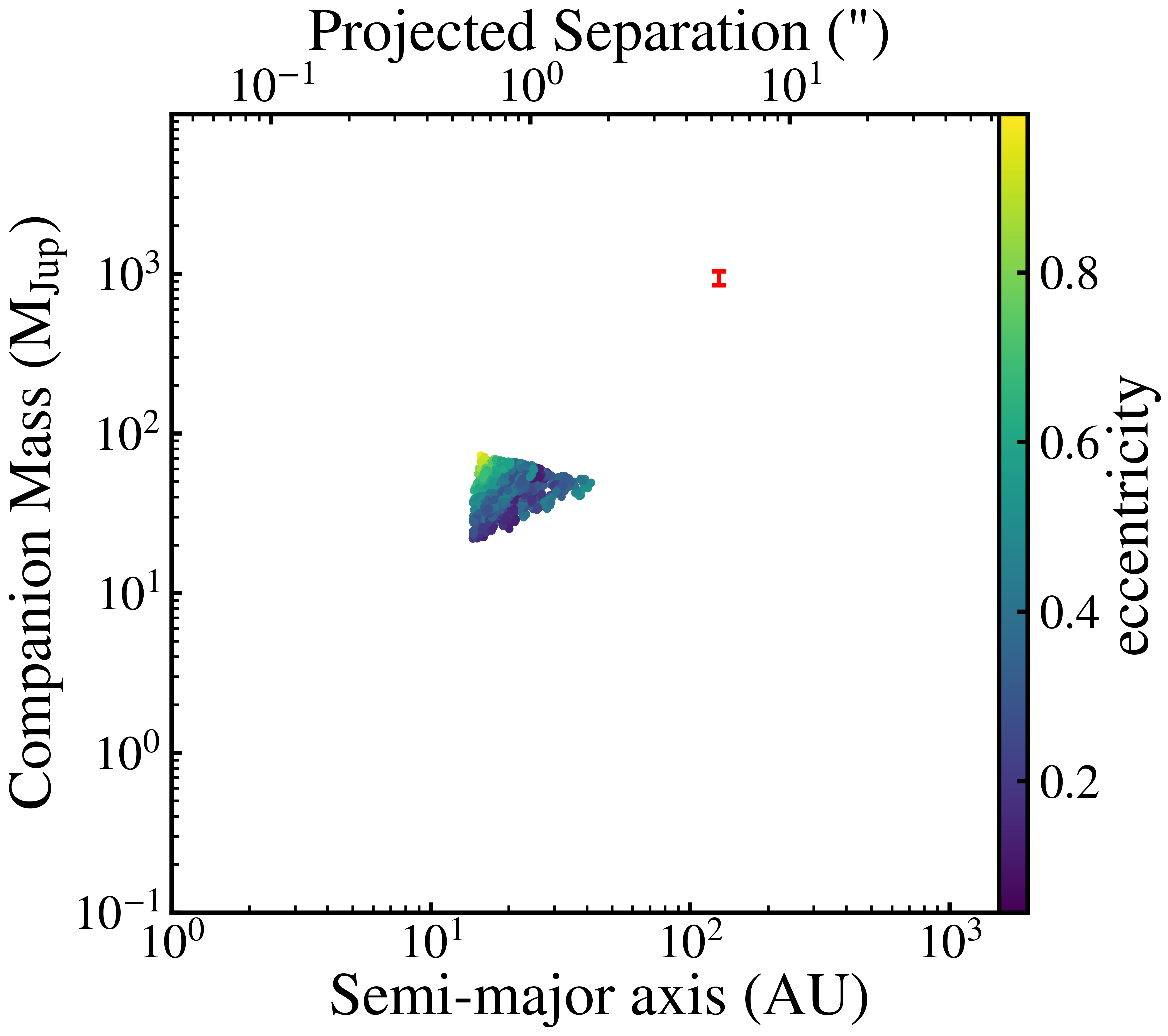}
    \end{minipage}
    
    \caption{Same as Figure \ref{fig:ethraid_octo_HIP18512} for HIP 45839. We use a vertical red line to indicate the range of possible masses, $0.81-0.99~\Msun$, for the imaged companion described in Section \ref{subsec:results_hip45839}, assuming that it is a bound white dwarf. If the imaged companion is not responsible for the measured accelerations, then the non-detection of any other sources in our Lick/Shane and Keck/NIRC2 imaging rules out stars and high-mass brown dwarfs ($M\sim80~\Mjup$) as the source of the accelerations. Our \texttt{ethraid} posteriors suggest that the RV and astrometric trends are consistent with planetary and brown dwarf models, with respective probabilities of $p$(planet)=$\hipBpplEthraid$ and $p$(BD)=$\hipBpBDEthraid$. Meanwhile, \texttt{octofitter} favors higher-mass models, giving $a=\hipBOctoA$ AU and $M=\hipBOctoM$ \Mjup and ruling out planetary models at high confidence: $p$(planet)=$\hipBpplOcto$, $p$(BD)$\sim\hipBpBDOcto$. \texttt{octofitter} provides tighter constraints in mass and separation, as expected from its incorporation of additional astrometric data products. We used the 17-year RV trend to set a minimum period of $\sim$35 years ($9.6$ AU).}
    \label{fig:ethraid_octo_HIP45839}
\end{figure*}

\begin{figure*}[t]
    \centering
    \begin{minipage}[b]{0.75\linewidth}
        \centering
        \includegraphics[width=\linewidth]{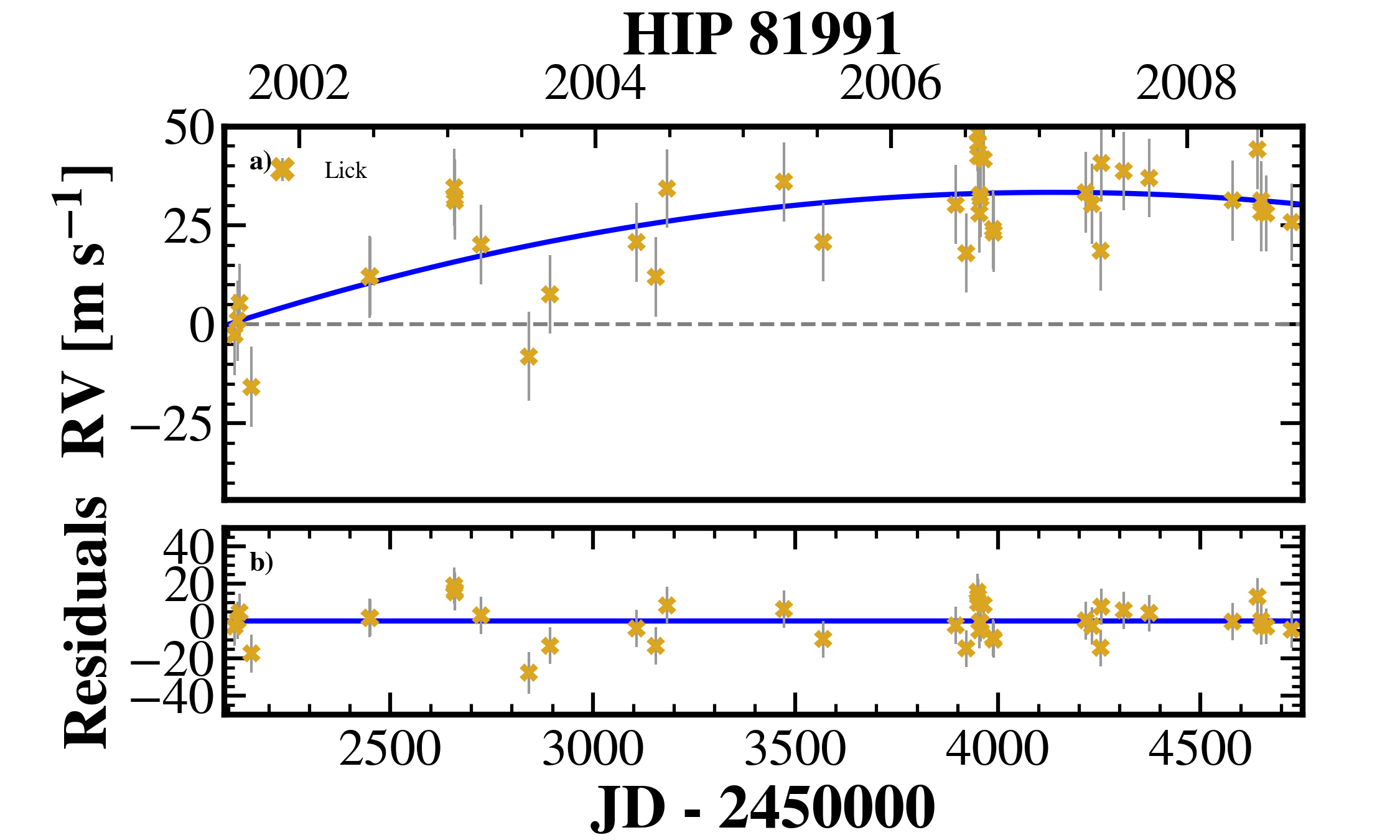}
    \end{minipage}
    \begin{minipage}[b]{0.95\linewidth}
        \centering
        \includegraphics[width=\linewidth]{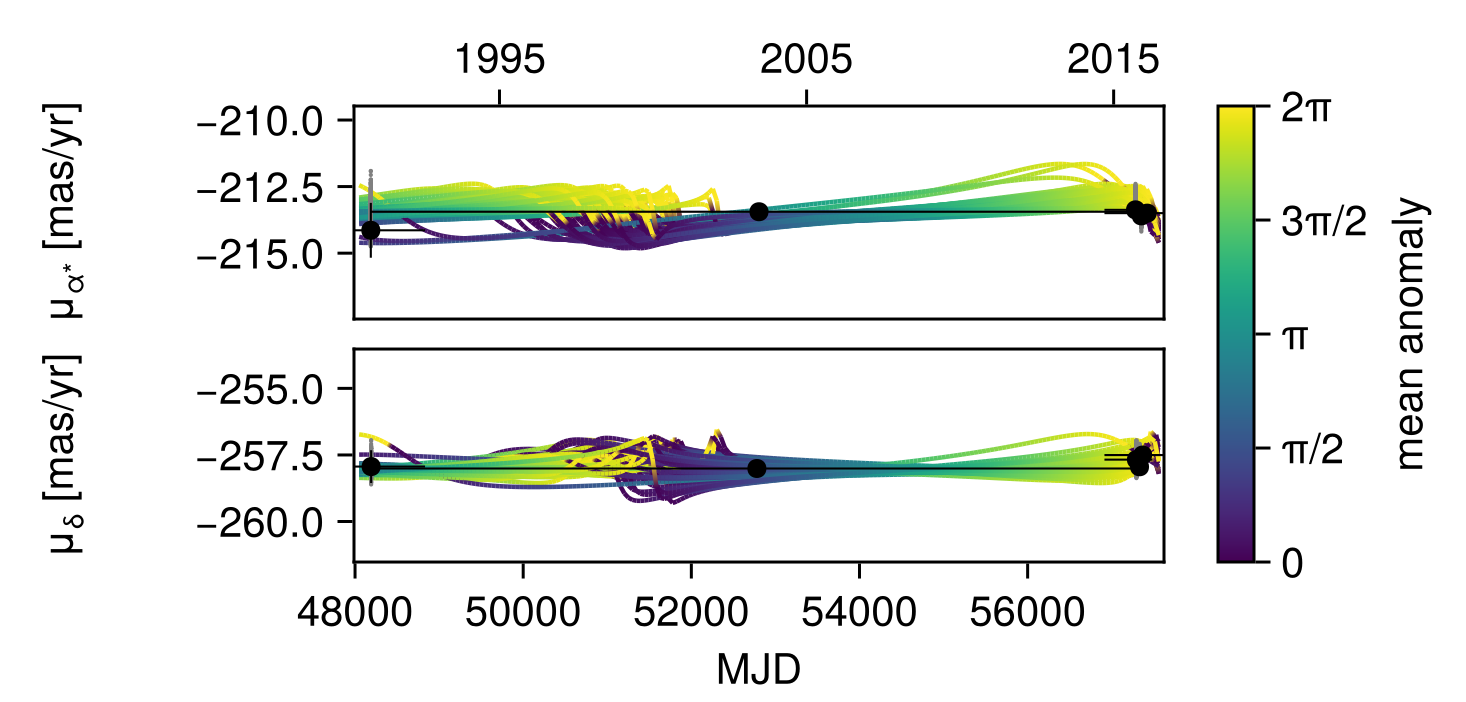}
    \end{minipage}
    
    \caption{Same as Figure \ref{fig:rv_astro_fits_HIP18512} for HIP 81991. We fit a trend with marginally significant (2.8$\sigma$) curvature to the RVs, and observe potential nonlinear variation in the proper motion fit, particularly in declination, which permits a wide range of feasible orbital models.}
    \label{fig:rv_astro_fits_HIP81991}
\end{figure*}

\begin{figure*}[t]
    \centering
    \begin{minipage}[b]{0.462\linewidth}
        \centering
        \includegraphics[width=\linewidth]{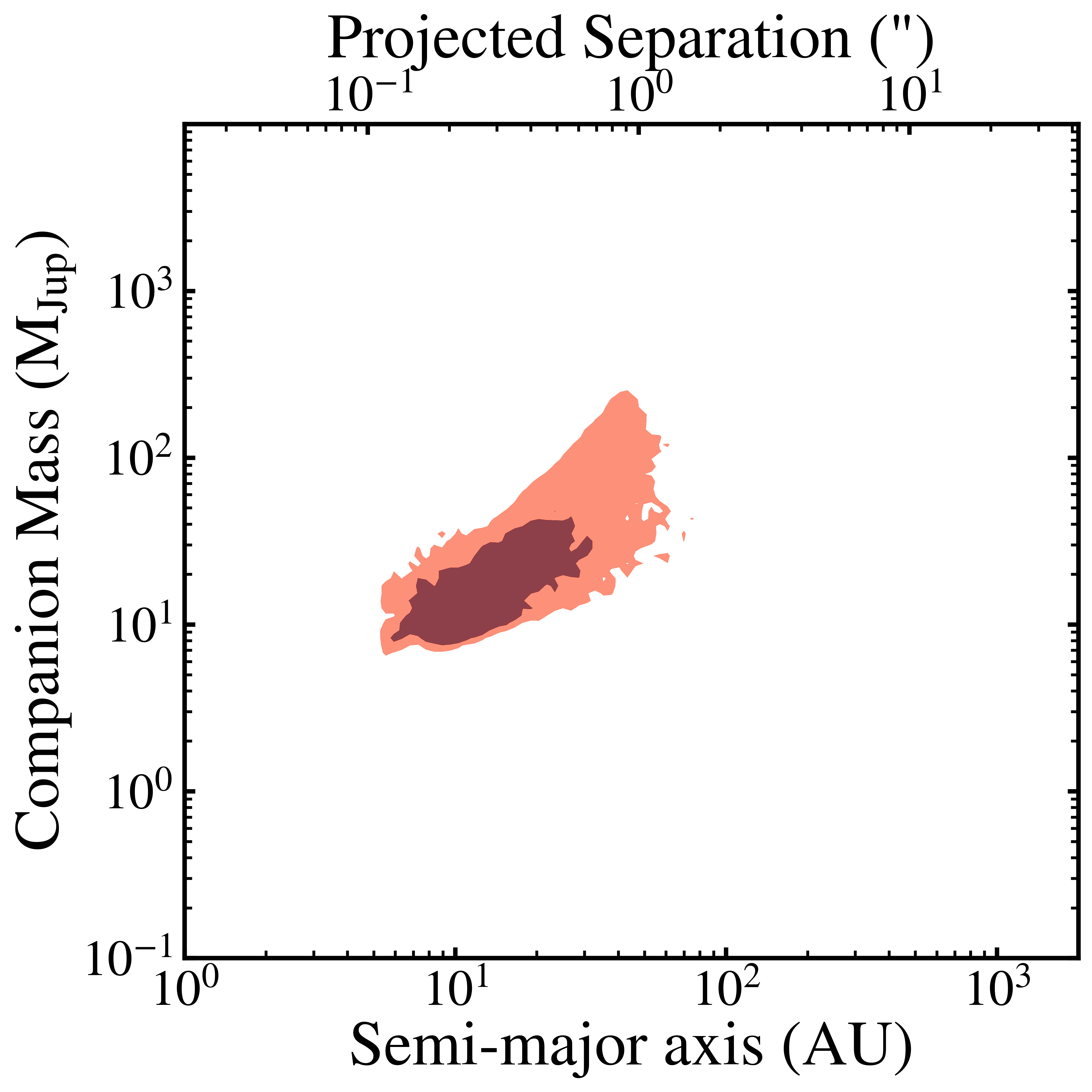}
    \end{minipage}
    \hfill
    \begin{minipage}[b]{0.52\linewidth}
        \centering
        \includegraphics[width=\linewidth]{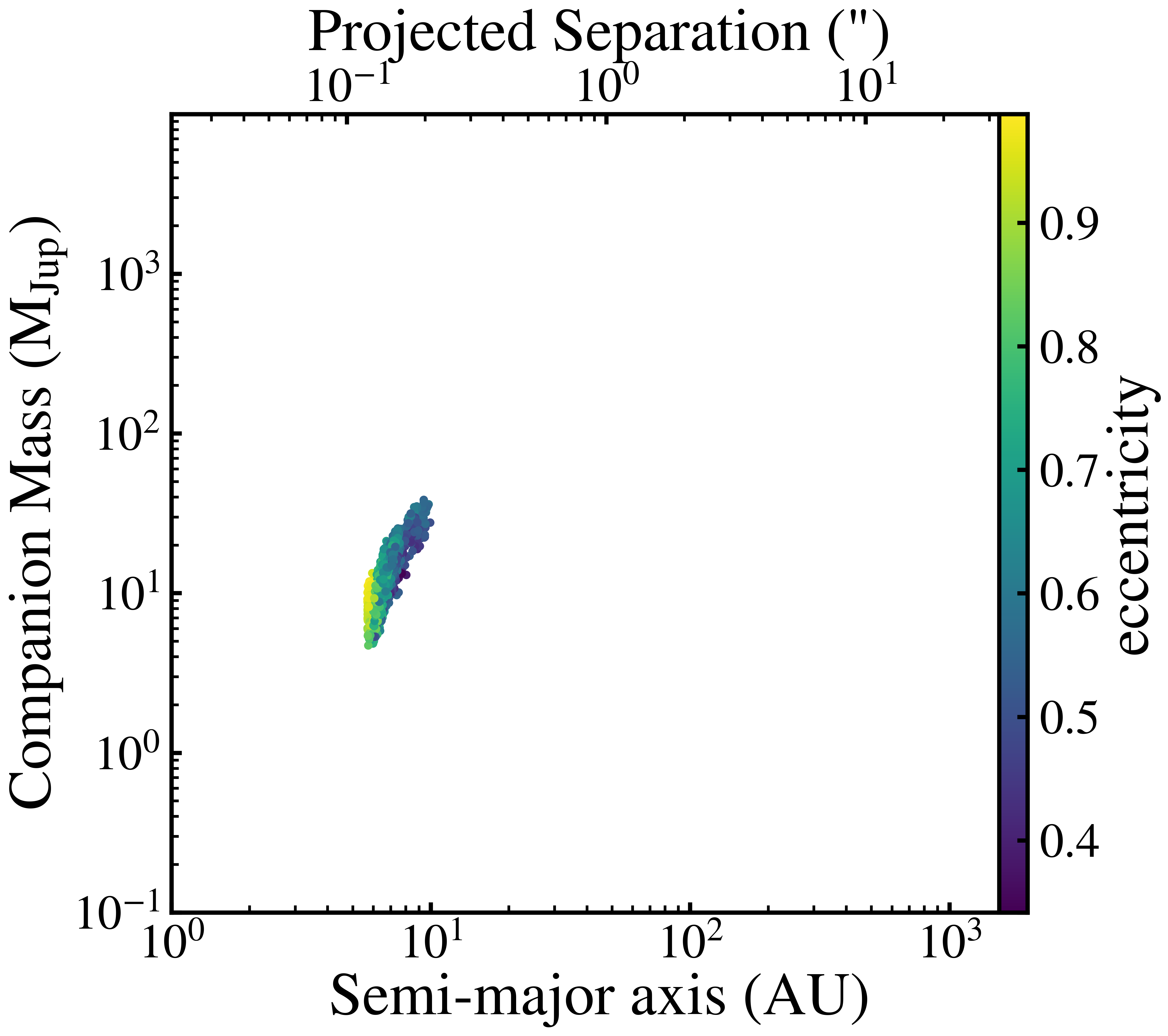}
    \end{minipage}

    \caption{Same as Figure \ref{fig:ethraid_octo_HIP18512} for HIP 81991. The high HGCA acceleration in this system rules out low-mass planetary models ($M<$a few \Mjup). Furthermore, RV curvature disfavors high-mass, high-separation models in our \texttt{ethraid} analysis, with Shane AO imaging ruling out the highest-mass models ($\gtrsim200~\Mjup$). RV curvature, along with additional short- and medium-baseline astrometric data, rules out high-mass models completely with \texttt{octofitter}. We find that the companion has $p$(planet)=$\hipCpplEthraid$, $p$(BD)=$\hipCpBDEthraid$, and $p$(star)=$\hipCpstarEthraid$ with \texttt{ethraid}. Using \texttt{octofitter}, we derive constraints of $a=\hipCOctoA$ AU and $M=\hipCOctoM$ \Mjup and find probabilities of $p$(planet)=$\hipCpplOcto$ and $p$(BD)=$\hipCpBDOcto$.}
    \label{fig:ethraid_octo_HIP81991}
\end{figure*}

The first code is \texttt{ethraid}, described in \cite{VanZandt2024}. \texttt{ethraid} uses a brute-force importance sampling approach to explore a broad range of candidate properties. Candidate masses and semi-major axes are drawn from log-uniform distributions and eccentricities and drawn from the beta distribution of \cite{Kipping2013}. Orbital angles and phases are drawn from uniform distributions, with the exception on inclination, which is drawn from a distribution uniform in $\cos(i)$. At each sample, \texttt{ethraid} computes the RV acceleration $\dot{\gamma}$ and jerk $\ddot{\gamma}$ at a reference time as well as proper motion between the $Hipparcos$ and $Gaia$ DR3 epochs and the proper motion within the $Gaia$ DR3 mission. These predicted values are compared against the observed value and samples are weighted according to their chi-squared likelihood. We generated $10^8$ samples per candidate and calculate the 2D  mass-separation joint posterior distributions. Figure \ref{fig:rv_astro_fits_HIP18512} shows our fits to the RV and astrometric data for HIP 18512, and we show our derived posterior distributions for this system in Figure \ref{fig:ethraid_octo_HIP18512}. We show analogous results for HIP 45839 in Figures \ref{fig:rv_astro_fits_HIP45839} and \ref{fig:ethraid_octo_HIP45839}, and for HIP 81991 in Figures \ref{fig:rv_astro_fits_HIP81991} and \ref{fig:ethraid_octo_HIP81991}.

The second code is \texttt{octofitter} \citep{Thompson2023}. We use \texttt{octofitter} v8.0 to perform full orbit fits using the non-reversible parallel tempered sampler \texttt{Pigeons.jl} \citep{Surjanovic2023}. We initialized \texttt{Pigeons} at eight sampling rounds, corresponding to $2^8=256$ retained posterior samples, and progressively increased the number of rounds to improve convergence. We found that for 13 rounds (8192 posterior samples), the split-$\hat{R}$ values \citep{Gelman2013} are close to unity (split-$\hat{R}$<1.004) and the change in log-evidence is stable for additional rounds of sampling ($\Delta \ln Z<0.5$ between rounds 12 and 13). The split-$\hat{R}$ statistic is similar to the Gelman-Rubin statistic $\hat{R}$, which compares the average variance within individual chains to the variance between independent chains. In the split-$\hat{R}$ formulation, each chain is divided in half and treated as two separate chains, making the statistic more sensitive to non-stationarity within chains. The fitted orbital parameters are the orbital period and companion mass (both sampled log-uniformly), eccentricity (sampled uniformly from 0 to 0.99), argument of periastron, inclination, longitude of ascending node, and the position angle $\theta$ at a reference epoch, which is chosen as BJD=2457389.0 (2016 January 1 12:00:00 UT). With \texttt{octofitter}, we jointly fit the absolute astrometry data from $Gaia$ and $Hipparcos$, as well as the RV data. The absolute astrometry includes the HGCA proper motion anomaly (\citealt{Brandt2021a}; between $Hipparcos$ and $Gaia$ DR3), calibrated $Gaia$ DR2-DR3 proper motions and scaled positional differences, $Gaia$ astrometric excess noise (AEN), and $Hipparcos$ intermediate astrometric data (IAD). The methodology and cross-calibration between $Gaia$ DR2 and DR3 are described in detail in \citet{Thompson2026}, while the $Gaia$ DR3 to $Hipparcos$ cross-calibration follows that of \citet{Brandt2021a}. We plot the \texttt{octofitter} orbit draws alongside our \texttt{ethraid} posteriors in Figures \ref{fig:ethraid_octo_HIP18512}, \ref{fig:ethraid_octo_HIP45839}, and \ref{fig:ethraid_octo_HIP81991}.

To compare the codes at a high level, \texttt{ethraid}, by construction, fully explores all relevant mass-separation space, but has low sampling efficiency, i.e. the vast majority of the draws have low likelihood. Although we modeled $10^8$ orbits per system, only $\sim10\%$ of these had non-zero likelihood values within machine precision. For each system we calculated the Kish effective sample size \citep{Kish1965}, defined as $\text{ESS}=\frac{(\sum_i w_i)^2}{\sum_i w_i^2}$ for sample weights $w_i$, to estimate the number of effectively independent samples in our simulations. We found ESS=500,000 for HIP 18512, 6,600 for HIP 45839, and 26,000 for HIP 81991. Thus for HIP 18512 we drew $\sim60$ times as many effective samples with \texttt{ethraid} as with \texttt{octofitter}, whereas for HIP 45839 and HIP 81991 the effective sample sizes were of the same order.

In contrast, \texttt{octofitter} explores the region of high posterior probability density with fewer samples, but care is required to ensure that all credible models are not missed due to difficult posterior topology. In general, the posterior distributions we calculate using \texttt{octofitter} are more constrained than those of \texttt{ethraid}, owing to the former's inclusion of additional astrometric data products spanning a wider range of time scales. We therefore adopt our \texttt{octofitter} results, and quote the \texttt{ethraid} results alongside them for comparison.

For each star, we sampled over the same mass and separation bounds with both \texttt{ethraid} and \texttt{octofitter}. We sampled masses between 1 \Mjup and the host star mass. We determined the minimum sampling separation as follows: we set a lower bound on the companion period at twice the system's current RV baseline, reasoning that no system's RVs show sufficient curvature to span more than half of one full orbit. We converted this period to a separation using Kepler's Third Law, and set a uniform upper separation bound of 600 AU. Table \ref{tab:priors} lists our priors for both codes. Although our \texttt{octofitter} results offer precise results for some of our systems, the derived posteriors are generally asymmetric. We quote the calculated mass and separation from \texttt{octofitter} as the peak value of the posterior distribution, with error bars spanning the 68\% highest density interval.

\begin{deluxetable}{cccc}
\tabletypesize{\footnotesize}
\tablecaption{Partial Orbit Results}
\label{tab:trends}

\tablehead{
    \colhead{Parameter} &
    \colhead{HIP 18512} &
    \colhead{HIP 45839} &
    \colhead{HIP 81991}
}

\startdata
$\dot{\gamma}$ (m/s/yr) & \hipAtrend & \hipBtrend & \hipCtrend \\
$\ddot{\gamma}$ (m/s/yr$^2$) & \hipAcurv & \hipBcurv & \hipCcurv \\
$\Delta \mu$ (mas/yr)   & \hipAdmu   & \hipBdmu   & \hipCdmu   \\
\hline
\multicolumn{4}{c}{\textbf{\texttt{octofitter}\tablenotemark{a}}} \\
\hline
$a$ (AU)     & --- & \hipBOctoA    & \hipCOctoA    \\
$M$ (\Mjup)    & ---   & \hipBOctoM    & \hipCOctoM    \\
$p$(pl)          & 0\%   & \hipBpplOcto    & \hipCpplOcto    \\
$p$(BD)          & 0\%    & \hipBpBDOcto    & \hipCpBDOcto    \\
$p$(star)        & 100\% & \hipBpstarOcto  & \hipCpstarOcto  \\
\hline
\multicolumn{4}{c}{\textbf{\texttt{ethraid}}} \\
\hline
$a$ range (AU)     & --- & \hipBEthraidA    & \hipCEthraidA    \\
$M$ range (\Mjup)    & ---   & \hipBEthraidM    & \hipCEthraidM    \\
$p$(pl)          & 0\%   & \hipBpplEthraid    & \hipCpplEthraid    \\
$p$(BD)          & 0\%    & \hipBpBDEthraid    & \hipCpBDEthraid    \\
$p$(star)        & 100\% & \hipBpstarEthraid  & \hipCpstarEthraid  \\
\hline
\enddata
\tablecomments{We report 68\% confidence mass and separation uncertainties for our \texttt{octofitter} results, and 95\% confidence intervals for \texttt{ethraid}. The quoted probabilities are based on a boundary of 13 \Mjup between planets and brown dwarfs, and 80 \Mjup between brown dwarfs and stars.}

\end{deluxetable}
\begin{deluxetable}{lcc}
\tabletypesize{\footnotesize}
\tablecaption{Priors on Partial Orbit Fits}
\label{tab:priors}

\tablehead{
    \colhead{Parameter} &
    \colhead{\texttt{ethraid}} &
    \colhead{\texttt{octofitter}}
}

\startdata
Semi-major axis $a$ & log-uniform &  --- \\
Period $P$ & --- &  log-uniform \\
Companion mass $M$ & log-uniform & log-uniform \\
eccentricity $e$ & $\mathcal{B}(0.867, 3.03)$ & $\mathcal{U}(0, 0.99)$\\
inclination $i$ & $\mathcal{U}(\cos i)$ & $\mathcal{U}(\cos i)$\\
Argument of periastron $\omega$ & uniform & uniform\\
Longitude of the ascending node $\Omega$ & uniform & uniform\\
Starting position at reference epoch & uniform & uniform\\
\enddata
\tablecomments{$\mathcal{B}(0.867, 3.03)$ is the two-parameter beta distribution with parameters derived by \cite{Kipping2013}, and $\mathcal{U}$ is a uniform distribution. \texttt{ethraid} uses the beginning of the $Hipparcos$ mission, 1989.5, as its reference epoch, while for \texttt{octofitter} we used BJD=2457389.0.}

\end{deluxetable}

\section{Results}
\label{sec:results}

\subsection{HIP 18512}
\label{subsec:results_hip18512}

The combined RV and astrometric accelerations of HIP 18512 are consistent with planetary, brown dwarf, and stellar companions. Follow-up AO imaging identified the previously known stellar companion to this star at a separation of 10\farcs87. \cite{An2025} performed a full orbital fit using RVs, HGCA astrometry, and relative astrometry from the WDS. They found that the binary companion has $M=0.552^{+0.064}_{-0.059}~\Msun$, $a=145^{+32}_{-14}$ AU, $i=140.8^{+7.3}_{-8.2}$$^{\circ}$, and $e=0.435^{+0.062}_{-0.110}$. The measured mass and separation are fully consistent with the observed RV and astrometric trends (see Figure \ref{fig:ethraid_octo_HIP18512}).


\subsection{HIP 45839}
\label{subsec:results_hip45839}

We detect a faint source ($K_s =$ 12.5 mag) at a separation of $5\farcs3$ in the NIRC2 imaging for this star (Figure \ref{fig:imaging}). Assuming the object is an intermediate-age ($\sim$5 Gyr) main sequence star comoving with the primary, its luminosity implies a mass of $\sim$80 \Mjup \citep{Baraffe2003}, insufficient to have caused the observed RV and astrometric accelerations at the projected separation of 128 AU. 

Another possibility is that the source is a white dwarf, in which case the observed brightness could be consistent with a higher mass and therefore with having caused the trend. To test this hypothesis, we used the MESA Isochrones and Stellar Tracks white dwarf cooling models (MIST; \citealt{Tremblay2011}; \citealt{Bauer2026}) to predict the expected brightness of a white dwarf companion and compared it to the observed companion luminosity. We sampled white dwarf parameters over a wide range of masses and cooling ages, finding that comoving white dwarfs between $0.81-0.99~\Msun$ and cooling ages of $11-400$ Myr would have brightnesses consistent with the observed $K$-band magnitude of 10.58. White dwarfs in this mass range would have progenitor masses of $\sim3.25-4.5~\Msun$ \citep{Cummings2018}, corresponding to main sequence lifetimes of 530 Myr and 230 Myr, respectively. Adding the simulated cooling ages gives total system ages of $\sim630-930$ Myr, a range that is in tension with the system age of $4.07^{+9.42}_{-2.84}~\mathrm{Gyr}$ reported by \cite{YeePetigura2017} at $\gtrsim1\sigma$. Due to this discrepancy, as well as the inherent scarcity of white dwarfs with $M>0.8~\Msun$ ($\sim7\%$, \citealt{Liebert2005}), we reason that a bound white dwarf is a possible but unlikely explanation for the trend in this system.

The astrometric acceleration of HIP 45839 provides another piece of evidence against the bound companion scenario. We observed the companion close to due North of the primary ($\theta = -4.57^{\circ}\pm 0.06^{\circ}$), but HIP 45839 is accelerating in nearly the opposite direction, $\theta_{\text{accel}}=194^{\circ}\pm8^{\circ}$, contrary to the expected behavior for a gravitationally bound system.

Assuming the imaged companion is not responsible for HIP 45839's acceleration, our imaging data allows us to rule out companions at 20 AU down to a mass limit of $\sim$80 \Mjup using ShARCS, and $\sim$60 \Mjup using NIRC2, implying a brown dwarf or planetary companion as the source of the astrometric and RV signals. We used \texttt{ethraid} to derive 95\% confidence intervals of $a=\hipBEthraidA$ AU and $M=\hipBEthraidM$ \Mjup, and classify the companion as follows: $p$(planet)=$\hipBpplEthraid$ and $p$(BD)=$\hipBpBDEthraid$, where we define ``planets" and ``brown dwarfs" as objects with $M<13~\Mjup$ and $M=13-80~\Mjup$, respectively. With \texttt{octofitter}, we found $a=\hipBOctoA$ AU and $M=\hipBOctoM$ \Mjup, and found that $p$(BD)$\sim\hipBpBDOcto$. We show our derived posteriors using both codes in Figure \ref{fig:ethraid_octo_HIP45839}. The median separation we found with \texttt{octofitter} corresponds to an orbital period of 90 years. Given that the current RV baseline of 17 years spans about 20\% of this period, continued RV monitoring may soon reveal significant curvature. 

Because HIP 45839 is moderately chromospherically active, an alternative explanation for the RV trend is long-term magnetic activity, which can increase the frequency and intensity of stellar spots, inducing apparent RV variation through suppression of convective blueshift \citep{Lindegren2003, Dumusque2011}. However, this explanation is disfavored by the astrometric data, which shows a trend consistent with the RVs despite being less affected by stellar activity. \cite{Lagrange2011} found that for a Sun-like star at a distance of 10 pc, the typical astrometric signal induced by stellar jitter has an amplitude of 0.2 $\mu$as, three orders of magnitude smaller than the proper motion anomaly of $\hipBdmu$ mas/yr observed for this star. Even for rapidly rotating stars, which tend to be more active \citep{Skumanich1972, Wright2011}, stellar jitter induces astrometric signatures on the order of 15 $\mu$as at 10 pc \citep{Sowmya2022}, roughly 4\% of the amplitude of HIP 45839's proper motion anomaly. The system distance of 24.15 pc would further diminish the effect of such signals.

\subsection{HIP 81991}
\label{subsec:results_hip81991}

The high proper motion anomaly exhibited by HIP 81991 ($0.51 \pm 0.03$ mas/yr) ruled out most low-mass planetary models ($M<$ a few $\Mjup$) as candidate causes for this system's astrometric and RV acceleration. Meanwhile, AO imaging and curvature in the RV data significantly reduced the fraction of long-period --- and therefore high-mass --- models consistent with the data (see Figure \ref{fig:ethraid_octo_HIP81991}).

Using the \texttt{ethraid} posterior we derived in Section \ref{sec:analysis}, we found $a=\hipCEthraidA$ AU and $M=\hipCEthraidM$ \Mjup at 95\% confidence. We probabilistically classify the companion as follows: $p$(planet)=$\hipCpplEthraid$, $p$(BD)=$\hipCpBDEthraid$, $p$(star)=$\hipCpstarEthraid$. By contrast, \texttt{octofitter} favors low-mass models: $p$(planet)=$\hipCpplOcto$, $p$(BD)=$\hipCpBDOcto$, $p$(star)=$\hipCpstarOcto$, notably ruling out stellar companions even before the inclusion of imaging data, using RVs and astrometry alone: $a=\hipCOctoA$ AU and $M=\hipCOctoM$ \Mjup. For this star, the $Gaia$ DR2-DR3 proper motion differences are significant at the $4.5\sigma$ level (calculated in a similar manner to HGCA astrometric accelerations), which is consistent with the curvature seen in the RV data and explains why \texttt{octofitter} favors shorter-period and lower-mass companion models than \texttt{ethraid}. 

HIP 81988 ($\rho=163''$, $\Delta$V$\sim$4) is bound to HIP 81991, as evidenced by its common proper motion. However, the star's wide separation is inconsistent with having produced the measured accelerations. As in the case of HIP 45839, the RV trend in this system is not likely to have arisen from stellar chromospheric activity due to the presence of a consistent astrometric proper motion anomaly, which itself has too large of an amplitude to be activity-induced.

\section{Conclusion}
\label{sec:conclusion}

$Gaia$ DR4 will precipitate a surge of exoplanet detections beginning in late 2026. The efficiency and reliability of these detections hinges on the preparation of effective tools and vetted target samples ahead of DR4's release. Here we have introduced GEODES, a survey designed to identify and characterize the most likely substellar companion hosts observed by $Gaia$. We highlighted three accelerating targets --- HIP 18512, HIP 45839, and HIP 81991 --- identified through our early efforts, and characterized the companion responsible for the acceleration in each system. Using two orbit-fitting codes, \texttt{ethraid} and \texttt{octofitter}, each designed to constrain companion orbital parameters by combining RVs, astrometry, and direct imaging, we recovered the stellar companion to HIP 18512, and ruled out stellar companions in the other two systems. We found that HIP 45839 hosts a brown dwarf ($a=\hipBOctoA$ AU and $M=\hipBOctoM$ \Mjup), and that HIP 81991's companion has a $\hipCpplOcto$ probability of being planetary, compared to a $\hipCpBDOcto$ probability of being a brown dwarf ($a=\hipCOctoA$ AU and $M=\hipCOctoM$ \Mjup).

Our results are representative of some of the expected future outcomes of our survey. Given the high fraction of stellar companions to Sun-like stars, we expect such companions to comprise many of our candidate planets. Thoroughly vetting our sample for stellar companions will allow us to commit more observational resources to characterizing truly substellar companions.

\section{Acknowledgments}
This publication makes use of The Data \& Analysis Center for Exoplanets (DACE), which is a facility based at the University of Geneva (CH) dedicated to extrasolar planets data visualisation, exchange and analysis. DACE is a platform of the Swiss National Centre of Competence in Research (NCCR) PlanetS, federating the Swiss expertise in Exoplanet research. The DACE platform is available at https://dace.unige.ch.This work has made use of data from the European Space Agency (ESA) mission {\it Gaia} (\url{https://www.cosmos.esa.int/gaia}), processed by the {\it Gaia} Data Processing and Analysis Consortium (DPAC, \url{https://www.cosmos.esa.int/web/gaia/dpac/consortium}). Funding for the DPAC has been provided by national institutions, in particular the institutions participating in the {\it Gaia} Multilateral Agreement. B.P.B. acknowledges support from the Alfred P. Sloan Foundation. J.W.X is thankful for support from the Heising-Simons Foundation 51 Pegasi b Fellowship (grant \#2025-5887). 


\software{\texttt{astropy} \citep{astropy:2018},
          \texttt{ethraid} \citep{VanZandt2024},
          \texttt{octofitter} \citep{Thompson2023},
          \texttt{RVSearch} \citep{Rosenthal2021},
          \texttt{radvel} \citep{Fulton2018},
          \texttt{pyKLIP} \citep{wangPyklipPsfSubtraction_2015},
          \texttt{KLIP} \citep{soummerDetectionCharacterizationExoplanets_2012},
          \texttt{rain} (\href{https://github.com/jsnguyen/rain}{https://github.com/jsnguyen/rain}),
          \texttt{ccdproc} \citep{craigAstropyCcdprocV130post1_2017},
          \texttt{emcee} \citep{foreman-mackeyEmceeMcmcHammer_2013},
          \texttt{numpy} \citep{numpy:2020}, 
          \texttt{scipy} \citep{scipy:2020}
          }

\bibliography{gaia_sub}
\bibliographystyle{aasjournal}


\end{document}